\documentstyle[12pt,aaspp4]{article}

\newcommand{\co}{{\rm $^{12}$CO}}
\newcommand{\cothree}{{\rm $^{13}$CO}}
\newcommand{\cn}{{\rm C$^0$}}
\newcommand{\cp}{{\rm C$^+$}}
\newcommand{\ci}{\mbox{\rm [\ion{C}{1}]}}

\newcommand{\cii}{\mbox{\rm [\ion{C}{2}]}}
\newcommand{\hi}{\mbox{\rm \ion{H}{1}}}
\newcommand{\hii}{\mbox{\rm \ion{H}{2}}}
\newcommand{\jone}{($J=1\rightarrow0$)}
\newcommand{\jtwo}{($J=2\rightarrow1$)}
\newcommand{\jthree}{($J=3\rightarrow2$)}
\newcommand{\jfour}{($J=4\rightarrow3$)}
\newcommand{\fsone}{($^3$P$_1\rightarrow^3$P$_0$)}
\newcommand{\fscii}{($^2$P$_{3/2}\rightarrow^2$P$_{1/2}$)}
\newcommand{\percmcu}{cm$^{-3}$}
\newcommand{\percmsq}{cm$^{-2}$}
\newcommand{\kmpers}{km~s$^{-1}$}
\newcommand{\Kkmpers}{K~km~s$^{-1}$}
\newcommand{\intunits}{erg~s$^{-1}$~cm$^{-2}$~sr$^{-1}$}
\newcommand{\Ico}{\mbox{\rm I$_{\rm CO}$}}
\newcommand{\Ici}{\mbox{\rm I$_{\rm [CI]}$}}
\newcommand{\Icii}{\mbox{\rm I$_{\rm [CII]}$}}
\newcommand{\citoco}{\Ici/\Ico}
\newcommand{\ciitoco}{\Icii/\Ico}

\newcommand{\av}{\mbox{\rm A$_{\rm V}$}}
\newcommand{\xuv}{\mbox{$\chi_{\rm uv}$}}

\begin{document}

\title{Carbon in the N159/N160 Complex of the Large Magellanic Cloud}

\author{Alberto D. Bolatto\altaffilmark{1}, James M. Jackson\altaffilmark{2},}
\affil{Institute for Astrophysical Research, Department of Astronomy} 
\affil{Boston University, 725 Commonwealth Ave., Boston MA 02215} 
\author{Frank P. Israel\altaffilmark{3},}
\affil{Sterrewacht Leiden, P.O. Box 9513, NL 2300-RA Leiden, The Netherlands} 
\author{Xiaolei Zhang\altaffilmark{4}}
\affil{Smithsonian Astrophysical Observatory, 60 Garden St. MS 78, 
Cambridge, MA 02138} 
\and 
\author{Sungeun Kim\altaffilmark{5}}  
\affil{University of Illinois at Urbana-Champaign, Department of Astronomy,
1002 West Green St., Urbana, IL 61801}

\altaffiltext{1}{bolatto@bu.edu}
\altaffiltext{2}{jackson@bu-ast.bu.edu}
\altaffiltext{3}{israel@strw.leidenuniv.nl}
\altaffiltext{4}{Currently at Raytheon ITSS and NASA GSFC, code 685, Greenbelt, MD 20771 (xzhang@specs1.gsfc.nasa.gov)}
\altaffiltext{5}{Currently at Smithsonian Astrophysical Observatory (skim@cfa.harvard.edu)}

\begin{abstract}
We present a study of carbon in N159/N160, an \hii\ region complex in
the low metallicity Large Magellanic Cloud. We have mapped this
region, which comprises four distinct molecular clouds spanning a wide
range of star-formation activity, in four transitions: \cothree\
\jone, \co\ \jtwo\ and \jfour, and \ci\ \fsone. Combining these data
with existing \cii\ \fscii\ observations provides a complete picture
of the predominant forms of carbon in the gas phase of the ISM. The
new CO \jtwo\ data show that the complex is immersed in an envelope of
extended, low level emission, undetected by previous \jone\ mapping
efforts.  The \co\ \jtwo/\jone\ ratio in this envelope is $\gtrsim3$,
a value consistent with optically thin CO emission. The envelope is
also relatively bright in \ci\ and \cii, and calculations show that it
is mostly photodissociated: it appears to be translucent ($\av<1$).
Neutral carbon emission in the complex unexpectedly peaks at the
quiescent southern cloud (N159S). In the northern portion of the map
(the N160 nebula), the \hii\ regions prominent in \cii\ correspond to
holes in the \ci\ distribution. Overall we find that, while the
\ciitoco\ ratio is enhanced with respect to similar complexes in the
Milky Way, the \citoco\ ratio appears to be similar or reduced.
\end{abstract}


\section{Introduction}
\label{intron159}
The Magellanic Clouds are especially interesting systems for the study
of the interstellar medium (ISM). Their low heavy element abundances
and low dust-to-gas ratios, coupled with their proximity and their
active star formation, make them ideal laboratories for understanding
how metallicity and strong radiation fields influence the composition
and structure of the ISM.  Furthermore, in several important aspects
they resemble primeval galaxies: they are morphologically irregular,
metal-poor, and very actively forming massive stars.  It is, however,
worthwhile to point out that the Magellanic Clouds are not pristine
systems. Their irregular morphologies refer mostly to the young
populations (massive stars, OB associations, and \hii\ regions),
probably influenced by recent encounters with the Milky Way: the
distribution of old stars and overall mass is not irregular
(Cioni, Habing \& Israel 2000). Nevertheless, because of a relatively
low time-averaged star formation rate the Magellanic Clouds are
considerably less developed that the Milky Way.  Therefore their study
will advance the understanding of the interaction between star
formation and the ISM in primordial systems.

The \hii\ regions in the N159/N160 complex were first cataloged by
Henize (1956) in a survey of emission line nebulae in the Large
Magellanic Cloud (Davies, Elliot, \& Meaburn 1976 give them the
numbers 271 and 284).  These nebulae are two of the brightest \hii\
regions in the immediate vicinity of 30 Doradus, located 
$\sim40\arcmin$ to its south. 

In this region the star formation activity appears to be progressing
from the (relatively) evolved starburst of 30 Doradus, where most of
the enshrouding molecular cloud has been dissipated (Johansson et
al. 1998), towards the quiescent southern CO arm region were little or
no star formation is currently taking place (Cohen et al. 1988; Kutner
et al. 1997).  Located at such a transitional place the N159/N160
complex is one of the best studied star-forming regions of the Large
Magellanic Cloud in molecular and atomic transitions (e.g., Johansson
et al. 1994; Israel et al. 1996; Stark et al. 1997a; Pak et
al. 1998; Johansson et al. 1998; Heikkil\"a, Johansson, \& Olofsson
1999). 

The N159/N160 complex features three distinct and spatially well
separated regions: 1) the northern region, chiefly associated with the
N160 nebula, where massive star formation is well evolved and the
parent clouds have been mostly, albeit not completely,
photodissociated and dissipated. 2) The central region, associated
with the N159 nebula, which is undergoing strong star formation
activity but still wrapped in molecular gas. This region includes two
giant molecular clouds (GMCs) known as N159-east and N159-west (N159E
and N159W respectively).  And, 3) the southern region, featuring the
molecular cloud N159-south (N159S).  This cloud is actually the
beginning of the 30 Doradus CO ridge region, a $\sim900$ pc
($\sim1^\circ$) long spur of molecular material extending southward of
the 30 Doradus nebula.  

The 30 Doradus CO ridge is the largest concentration of CO in the LMC
(Cohen et al. 1988). Because the ridge is located on the leading edge
of the LMC's movement through the hot Galactic halo (Mathewson \& Ford
1984), its formation has been attributed to ram pressure compression
of the Magellanic interstellar medium (de Boer et al. 1998; Kim et
al. 1998).  The entire ridge region (including N159S) is quiescent,
with little or no star formation activity as evidenced by its faint
far-infrared (FIR), H$\alpha$ (Davies et al. 1976), and \cii\ emission
(Mochizuki et al. 1994).

Because dust is the dominant source of UV extinction, the low
dust-to-gas ratio of the metal-poor ISM allows the UV radiation to
penetrate more deeply into the molecular material thereby causing
widespread photodissociation and photoionization. The regions, at the
surfaces of clouds, where the UV radiation dominates the heating and
chemistry of the ISM are known as photon-dominated or
photodissociation regions (PDRs). In these regions the UV radiation
dissociates most molecular species, but leaves H$_2$ relatively intact
because of its strong self-shielding (Abgrall et al. 1992; Pak et
al. 1998). Therefore in metal-poor systems an increasingly large
fraction of the molecular gas is expected to be associated with \cn\
and \cp, rather than CO. Observations of CO together with its
photodissociation and photoionization products (\cn\ and \cp) are
the key to understand the interplay between the radiation field and
the chemical and physical structure of molecular clouds in low
metallicity systems (e.g., Bolatto, Jackson, \& Ingalls 1999; Bolatto
et al. 2000). This paper presents some of these observations.

What is the dust-to-gas ratio in the LMC and how does it affect its
PDRs?  Several arguments suggest that the dust-to-gas ratio of a
galaxy is proportional to its metallicity $Z$ (Franco \& Cox 1986),
but recent studies find a $Z^2$ dependence (Lisenfeld \& Ferrara
1998).  Independently of this, the dust-to-gas ratios measured in the
Magellanic Clouds are $\sim1/4$ and $\sim1/8$ of the Galactic value
(LMC and SMC respectively, Koornneef 1982; Bouchet et al. 1985), while
their metallicities (measured through the oxygen abundance) are
$Z_{\rm LMC}\sim Z_{\rm Orion}/2$ and $Z_{\rm SMC}\sim Z_{\rm
Orion}/5$.  Because of this relative lack of dust, to achieve the same
visual extinction \av\ in the LMC requires $\sim4$ times larger gas
columns than in the Milky Way. Therefore, PDRs are expected to be on
average $\sim4$ times more extended in the LMC, while the CO cores
will be 4 times smaller (e.g., Israel 1997; Pak et al. 1998).

Observations suggest that CO cores for clouds of similar virial mass
are indeed smaller in the LMC than in the Milky Way, and most CO is
photodissociated.  The CO \jone\ emission from LMC clouds is several
times fainter than in the Milky Way (Cohen et al. 1988; Israel et
al. 1993; Rubio, Lequeux, \& Boulanger 1993; Kutner et
al. 1997). Moreover, the \ciitoco\ ratio in the LMC (which measures
the fraction of carbon in PDRs) appears to be
considerably enhanced with respect to the Milky Way (Mochizuki et
al. 1994; Poglitsch et al. 1995; Israel et al. 1996).  

\section{Observations}
\label{observations}
In this study we discuss four new datasets on the N159/N160 
region: \cothree\ \jone\ at $\nu\simeq110.2020$ GHz, 
\co\ \jtwo\ at $\nu\simeq230.5439$ GHz (Fig. \ref{sestmaps}), 
\co\ \jfour\ at $\nu\simeq461.0878$ GHz (Fig. \ref{co43map}), 
and \ci\ \fsone\ at $\nu\simeq492.1607$ GHz (Fig. \ref{cmap}).
The first two were acquired using the Swedish-ESO Submillimeter 
Telescope (SEST) at La Silla, Chile, during May 1999. The latter two 
were made with the 1.7 m Antarctic Submillimeter Telescope and Remote
Observatory (AST/RO, Stark et al. 1997b) at the Amundsen-Scott South Pole 
base, during the austral winter of 1998. Assuming an LMC 
distance of 52 kpc ($D.M.\simeq18.58$; Panagia, Gilmozzi, \& Kirshner 1997), 
the ``plate scale'' for these observations is $\approx15$ pc per arcminute.

To complement these millimeter and submillimeter-wave datasets, we
also use 21 cm \hi\ and radio continuum data from the ATCA (Australia
Telescope Compact Array) aperture synthesis survey of the LMC (Kim et
al. 1998). Approximately three pointing centers were included in this
map of the N159/N160 region. The zero-spacing flux, measured using the
Parkes telescope, has been added to the ATCA data to obtain reliable
column density estimates.

The \cothree\ \jone\ transition is excited in almost all conditions
found in molecular clouds (gas kinetic temperature $T_{kin}\gtrsim5$~K
and critical density $n_{cr}\sim1.7\times10^3$~cm$^{-3}$) and is
generally used as a tracer of gas column density, since it is
optically thinner than its isotopomer \co. The \co\ \jtwo\ transition
requires slightly warmer and denser gas to be excited
($T_{kin}\gtrsim11$~K, $n_{cr}\sim1.1\times10^4$~cm$^{-3}$), but these
conditions prevail in all except the coldest and most diffuse molecular
clouds. The excitation of the \co\ \jfour\ transition requires much
warmer and denser gas ($T_{kin}\gtrsim55$~K, $n_{cr}\sim10^5$
cm$^{-3}$). Thus \co\ \jfour\ is excited only in the dense clumps and in
the presence of heating sources. Finally, the \ci\ \fsone\ transition
is excited in very general conditions ($T_{kin}\gtrsim23$ K,
$n_{cr}\sim470$ cm$^{-3}$). Because \cn\ is understood to be chiefly
produced by the photodissociation of CO, this transition is thought to
trace the warm PDR that forms in the surfaces of molecular clumps
exposed to UV (we will, however, explore other alternatives later).
The precise values of the critical density in all the above cases
depend on the actual magnitude of radiative trapping effects, the
ratio of ortho- and para-H$_2$, and the kinetic temperature of the
collisional partners, and are thus only approximate.

\subsection{Millimeter Data}
The bulk of the SEST data has system temperatures in the range
220---350~K at 230 GHz and $T_{sys}\sim$190---250~K at 110 GHz.  The
maps were acquired during 1999 May, using the facility IRAM 115/230
GHz dual SIS receiver. The two frequencies were fed into the split
high resolution acousto-optical spectrometer (AOS) backend. The
resulting bandpasses ($\sim50$ \kmpers\ for 230 GHz and $\sim100$
\kmpers\ for 110 GHz) are more than adequate for the relatively narrow
lines and small velocity variation inside N159/N160. The spectra were
acquired in double beam switching mode, chopping 11.5\arcmin\ in
azimuth. A few of the spectra showed contamination by emission in the
off position, and these positions were reobserved at a different
parallactic angle.  At these frequencies the telescope has
HPBW(230)$\approx23\arcsec$ and HPBW(110)$\approx48\arcsec$, and the
main beam efficiency is $\eta_{mb}(230)\approx0.50$ and
$\eta_{mb}(110)\approx0.70$.  

The map was initially observed on a 24\arcsec\ grid (i.e.,
approximately full-beam spacing at 230 GHz) and the brightest emitting
regions were afterwards filled in on a 12\arcsec\ grid (half-beam
spacing). Two regions selected from the portion of the map showing low
level emission were reobserved several times simultaneously in \co\
\jone\ and \jtwo\ (boxes in Fig. \ref{sestmaps}). Pointing and focus
on a nearby SiO maser were performed before each observing
session. The SEST pointing accuracy is 3\arcsec\ rms.

\subsection{Submillimeter Data}
The \co\ \jfour\ AST/RO observations were made with a system
temperature $T_{sys}\sim3500$ K. They were acquired on 15 May 1998
using the lower frequency side of AST/RO's 460/810~GHz dual SIS
waveguide receiver. The \ci\ \fsone\ observations had a system
temperature $T_{sys}\sim1700$ K. The region was observed twice, on 1
July and 10 August of 1998, using the AST/RO SIS quasi-optical
receiver.  The backend in both cases was the 2048 channel low
resolution AOS providing a spectral resolution $\Delta v\sim0.4$
\kmpers\ over a bandpass of $\sim350$ \kmpers.  The spectra were
acquired in position switching mode, chopping 15\arcmin\ in azimuth
(which is the same as R.A. at the pole).  

The telescope was later
determined to have HPBW(461)$\sim3.4\arcmin\pm0.3$ and
HPBW(492)$\sim3.8\arcmin\pm0.3$, using scans across the limb of the
full Moon. The difference between the measured beams and the
diffraction limited beam size for the telescope was caused by a
2\arcdeg\ angular misalignment of the tertiary. No successful beam
maps were acquired during the season, however, and the precise beam
shape for either receiver is not actually known. Some of the source
elongation manifested in the peaks of Fig. \ref{co43map}, for example,
may not be intrinsic but caused by a distorted beam shape in that
particular receiver.  

The forward efficiency for AST/RO was determined from skydips to be
70\%, and it is assumed to be identical to the main beam efficiency as
the telescope is an off-axis, unblocked aperture dish.  Thus we assume
$\eta_{mb}\approx0.7$ for 461 and 492~GHz. The maps were observed on a
30\arcsec\ grid, considerably oversampling the beam. A few of the
spectra were contaminated by strong noise spikes that occasionally
occur in two specific regions of the passband, possibly associated
with electrical cross-talk between digital and analog signals in the
AOS. These problematic spectra were discarded.  Some grid points are
therefore missing, but because of the oversampling this does not
compromise the quality of the maps.  The sensitivity of the \ci\ map
is $1\sigma\approx0.1$ \Kkmpers\ in the central area, but somewhat
worse at the edges because there the spatial averaging cannot take
advantage of the heavy beam oversampling.

AST/RO's receivers sit on a static optical bench, located in the warm
room below the dish, that does not track with the telescope.
As a result any optical misalignment translates into precession of the
receiver beam in the sky, a problem for faint sources that require
long integrations.  AST/RO's radio pointing was carefully controlled
during the 1998 season by repeatedly observing a set of compact line
emitting sources at different elevations and fitting a model that
compensates for elevation-dependent pointing offsets and beam
precession. Thus we are confident that the pointing accuracy during
our observations was better than $\sim30\arcsec$ rms.

\section{Discussion}

\subsection{The Molecular ISM as Revealed by CO Observations}
\label{codiscussion}

The CO dataset is composed of new \cothree\ \jone, \co\ \jtwo, and
\co\ \jfour\ maps, as well as a preexisting \co\ \jone\ map obtained
by the SEST Magellanic Clouds key program.  Figure \ref{sestmaps}
shows the \co\ \jtwo\ and \cothree\ \jone\ SEST data.  The four main
molecular peaks (Table \ref{tabpos}) are apparent in both transitions:
N160 to the north is a relatively weak and elongated cloud, N159W is
the strongest peak in the complex, N159E appears to break up into
three clumps at this resolution, and N159S is a triangular cloud with
the smallest \jtwo/\jone\ intensity ratio in the complex. The cloud at
offset $\sim[+2\arcmin,+5\arcmin]$ is another member of the complex, 
which we will refer to as N160E. This cloud has very narrow lines
(FWHM$\sim 2.6$ \kmpers).  In addition to the bridge connecting 
N159S with the northern portion of the complex, there is a wealth of
extended, faint emission throughout the mapped region.  It is
important to stress this point: even below the 2 \Kkmpers\ contours
essentially every position in this map shows a significant CO \jtwo\
line if neighboring spectra are averaged. We will investigate the
character of this extended, low level emission in \S\ref{translucentCO}.

The overall correspondence of these data with the published \co\
\jone\ map (Johansson et al. 1994; Johansson et al. 1998) is extremely
good, although some differences are apparent in the relative
intensities and positions of the individual peaks.  The shift in
position is most noticeable for N159S, which at 1.3 mm appears to be
$\sim21\arcsec$ eastward of its nominal 2.6 mm position.  This offset
is most probably not real, but caused by undersampling.  These regions
were mapped on a 30\arcsec\ grid at \jone, while the SEST beam size at
115 GHz is 45\arcsec\ (Johansson et al. 1998). Our own \jtwo\ data are
only sampled every 230 GHz beam (24\arcsec\ grid) for N159S (N159E,
N159W and N160 were observed on a 12\arcsec\ grid).  The CO \jtwo\
peak in N159S falls almost precisely halfway between the \jone\
samplings, and its position agrees very well with the fully sampled
\cothree\ \jone\ map. This suggests that the \jtwo\ position for this
core may be better than the \jone\ location.  Excitation gradients
within N159S may also explain this positional discrepancy.
Fortunately, the shifts in the position of the peaks are much smaller
for the remaining clouds. To maintain consistency with previous work
(e.g., Heikkil\"a et al. 1999) we use the CO \jone\ positions
tabulated in Table \ref{tabpos}. Because the submillimeter data have
much lower angular resolution the precise position of the peaks will
not greatly affect the ratios in Table \ref{tabsubmm}.

\subsubsection{Physical Conditions}

\begin{figure}
\plotone{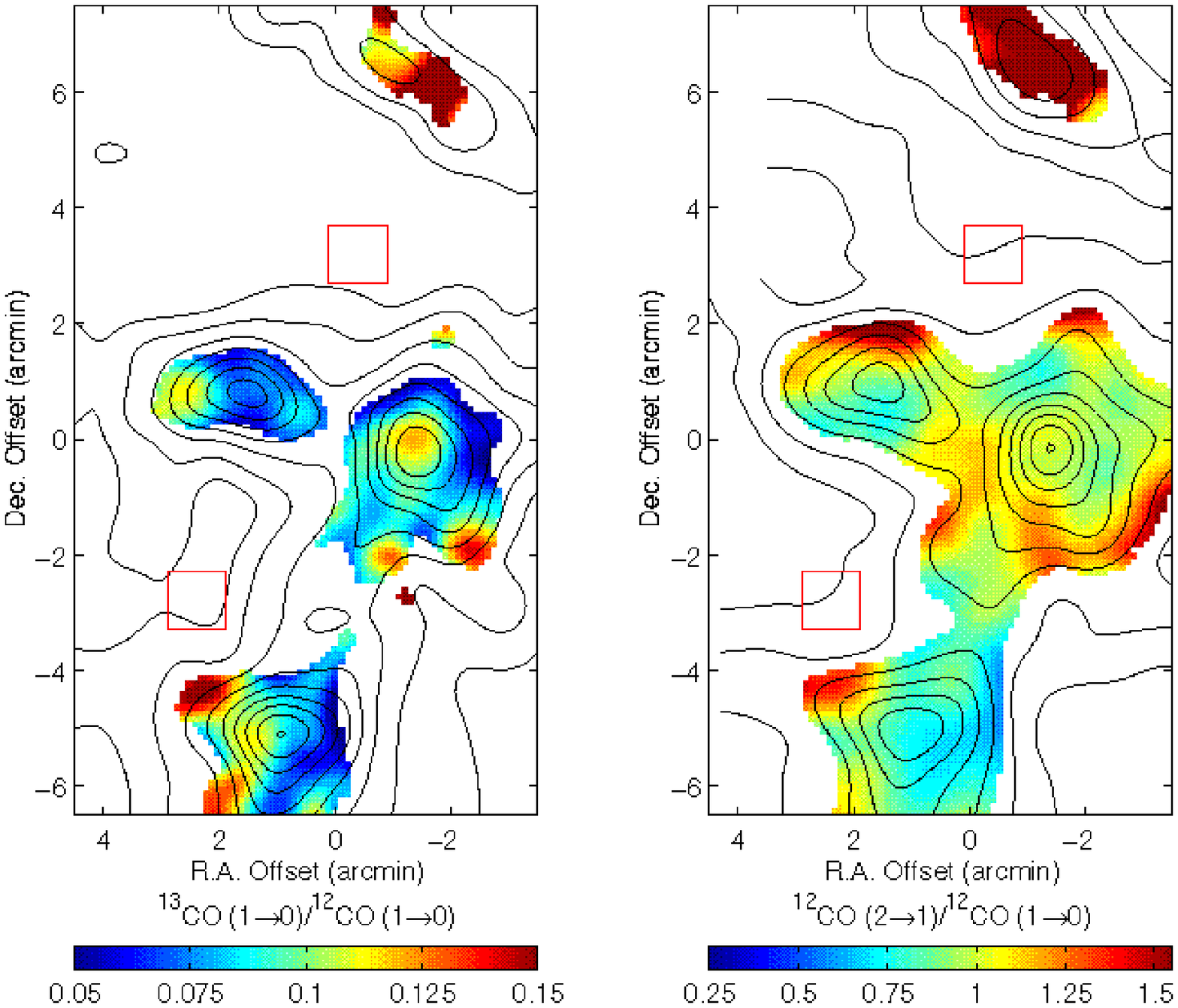} 
\figcaption[fig2.eps]{Comparison of \co\ \jtwo\ and \cothree\ \jone\
emission with published \co\ \jone\ data (Johansson et al. 1998).  All
maps have been Gaussian smoothed to a common resolution of 60\arcsec.
The contour levels are the same in both maps (2, 5, 10, 15, \ldots, 40
\Kkmpers).  Note that the \co\ \jone\ map was acquired by frequency
switching, and required an elaborate baseline subtraction
scheme. Therefore the lowest contour is very unreliable.  {\em (Left)}
Main beam integrated intensity \co\ \jone\ contour data (Johansson et
al. 1998) overlaid on the \cothree/\co\ \jone\ ratio image. Blue
denotes low \cothree/\co\ ratios and therefore optically thinner CO
emission, whereas red shows high ratios and optically thicker
emission.  {\em (Right)} Main beam integrated intensity \co\ \jtwo\
contour data overlaid on the \co\ \jtwo/\jone\ ratio image.  Blue
signifies low \jtwo/\jone\ ratios, and therefore possibly cooler (less
dense) gas; red zones are warmer (denser).
\label{coratios}}
\end{figure}

What are the physical conditions in the different clouds of the
complex?  Figure \ref{coratios} compares the \co\ \jone, \jtwo, and
the \cothree\ \jone\ observations. 
All three datasets have been spatially smoothed
to a common 60\arcsec\ resolution, and the results for the regions of
interest are summarized in Table \ref{tabmm}.  The \cothree/\co\
integrated intensity ratio is commonly used to estimate column
density. The \co\ \jtwo/\jone\ line ratio is sensitive to density in
the low density, optically thick limit, and to temperature for high
densities (e.g., Kaufman et al. 1999).  In the optically thin regime
this ratio is sensitive to temperature, assuming that the emission is
thermalized (for an exhaustive discussion of CO excitation see,
however, Warin, Benayoun, \& Viala 1996). 
 
The centers of the N160 and N159W clouds have the highest
\cothree/\co\ \jone\ ratios and therefore the largest opacities, while
at N159E this ratio is smaller by a factor of $\sim2$. Conversely,
N160 has the largest \co\ \jtwo/\jone\ ratio and therefore appears to
be the hottest, while N159S is the coldest cloud in the complex.
These conclusions agree broadly with a much more detailed multiline
excitation analysis performed toward the main molecular peaks in the
complex by Heikkil\"a et al. (1999), who find kinetic temperatures of
20, 25 and 10 K for N159W, N160 and N159S respectively.

\begin{figure}
\plotone{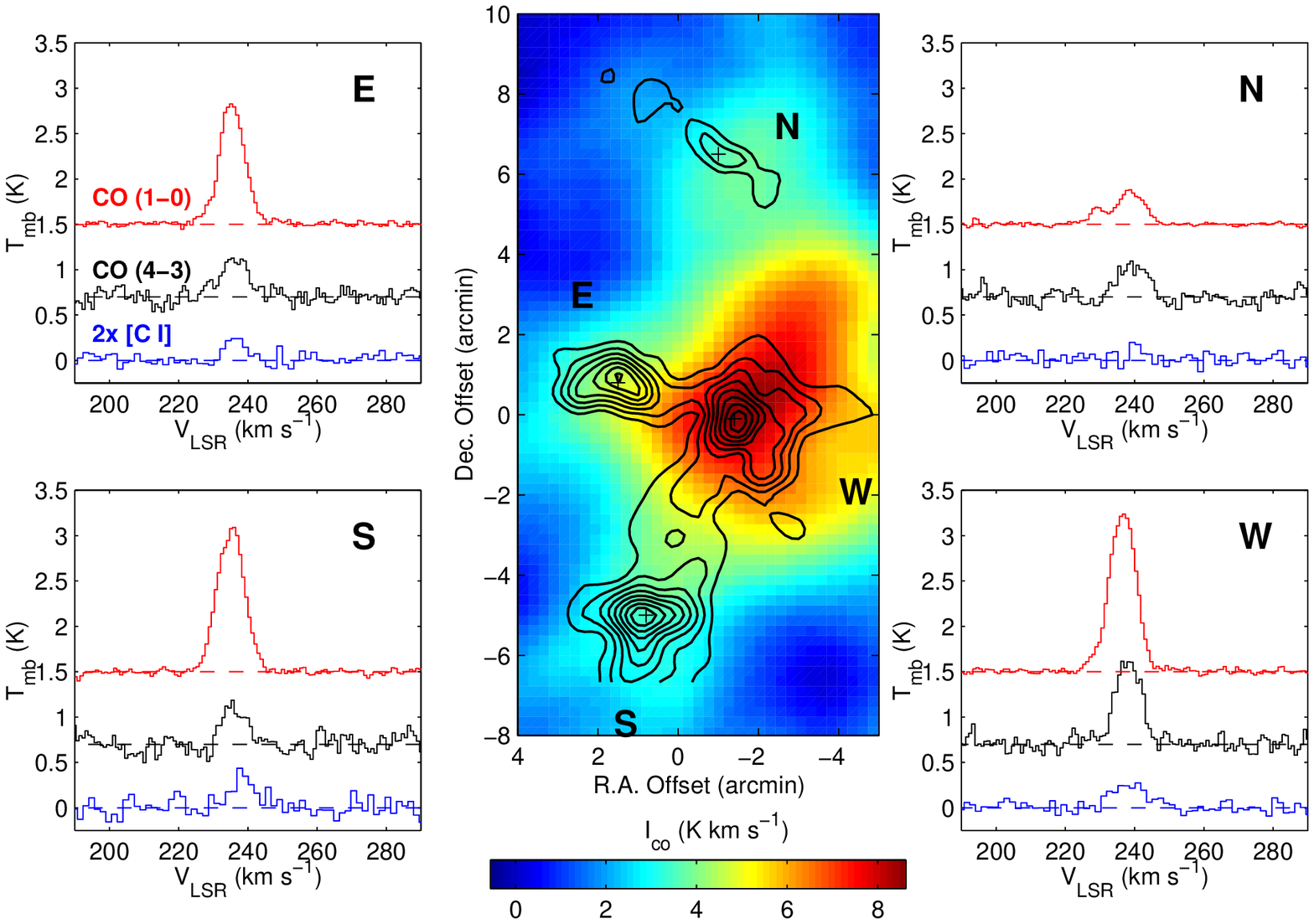} \figcaption[fig4.eps]{Color image of the \co\
\jfour\ integrated intensity map with \co\ \jone\ data overlaid. The
\jfour\ data has been smoothed to a 4\arcmin\ spatial resolution.
Marked by crosses are the positions of the four main molecular peaks
in the region.  Their corresponding \co\ \jfour\ (black) and \ci\
(blue) spectra are shown on the side (the \ci\ spectra have been
scaled up by a factor of two) For purposes of comparison, shown in red
are the \co\ \jone\ spectra (Johansson et al. 1998) convolved to the
same angular resolution (4\arcmin\ HPBW). The relative enhancement and
deficit of \jfour\ in the N160 (labeled N) and N159S (labeled S)
clouds respectively are apparent.
\label{co43map}}
\end{figure}

The distribution of the warm and dense gas in the complex is
elucidated by Fig. \ref{co43map}. Recall that the \co\ \jfour\
transition requires $T\gtrsim50$ K and $n\sim10^5$ cm$^{-3}$ to be
excited (\S \ref{observations}). The angular resolution of the \jfour\
observations is $\sim4\arcmin$, therefore all of the clouds are beam
diluted and it is necessary to consider their beam-filling fractions
to understand their relative intensities. All else being equal, larger
clouds would appear brighter in Fig. \ref{co43map}.  The peak of the
CO \jfour\ map is N159W, due to a combination of intense emission and
beam filling.  Although there is an extension of \jfour\ emission in
the southward direction, a peak at the position of N159S is most
noticeably absent. Because N159W and N159S have similar sizes and
intensities in the CO \jone\ transition, this implies that the CO
\jfour/\jone\ ratio is much smaller for N159S.  

Conversely, there is a hint of an emission peak at the position of
N160, a small cloud in \co\ \jone. The presence of this peak suggests
that the CO \jfour/\jone\ ratio for N160 is large. These results for
the ratios are confirmed by convolving the CO \jone\ map to the 
resolution of the \jfour\ data (Table \ref{tabsubmm}).  The CO
\jfour/\jone\ ratio strongly depends on temperature and density, but
the analysis performed by Heikkil\"a et al. (1999) shows that both
clouds have volume densities above the \jfour\ transition critical
density.  Therefore the deficit of \jfour\ emission in N159S is mostly
due to the much lower temperature of the southern cloud. In the
optically thick, LTE limit the observed CO \jfour/\jone\ ratio
corresponds to $T_{kin}\sim8$~K. This low temperature is probably due 
to the absence of star formation activity in N159S.

Figure \ref{heating} reveals the sources of heating and
photodissociating radiation in the N159/N160 complex.  The optical
N159 nebula is centered between the N159E and N159W clouds. Portions
of these clouds are in the foreground of the nebula and show
themselves as obscurations against the bright background. The N160
cloud, however, appears to be behind its nebula. Part of the N160
nebulosity fills in the gap between the N160 and N160E molecular
clouds, and is neatly delineated by them.  N159S is seen here only as
a subtle obscuration against the field stars, with no associated
nebulosity. While all the other clouds have apparent embedded sources
visible as peaks in the 60 $\mu$m IRAS HIRES picture, N159S appears
completely devoid of such sources. This lack of 60 $\mu$m emission
confirms the complete absence of any massive star-forming activity in
the southern cloud.

\subsubsection{The Nature of the Extended CO Emission}
\label{translucentCO}

\begin{figure}
\plotone{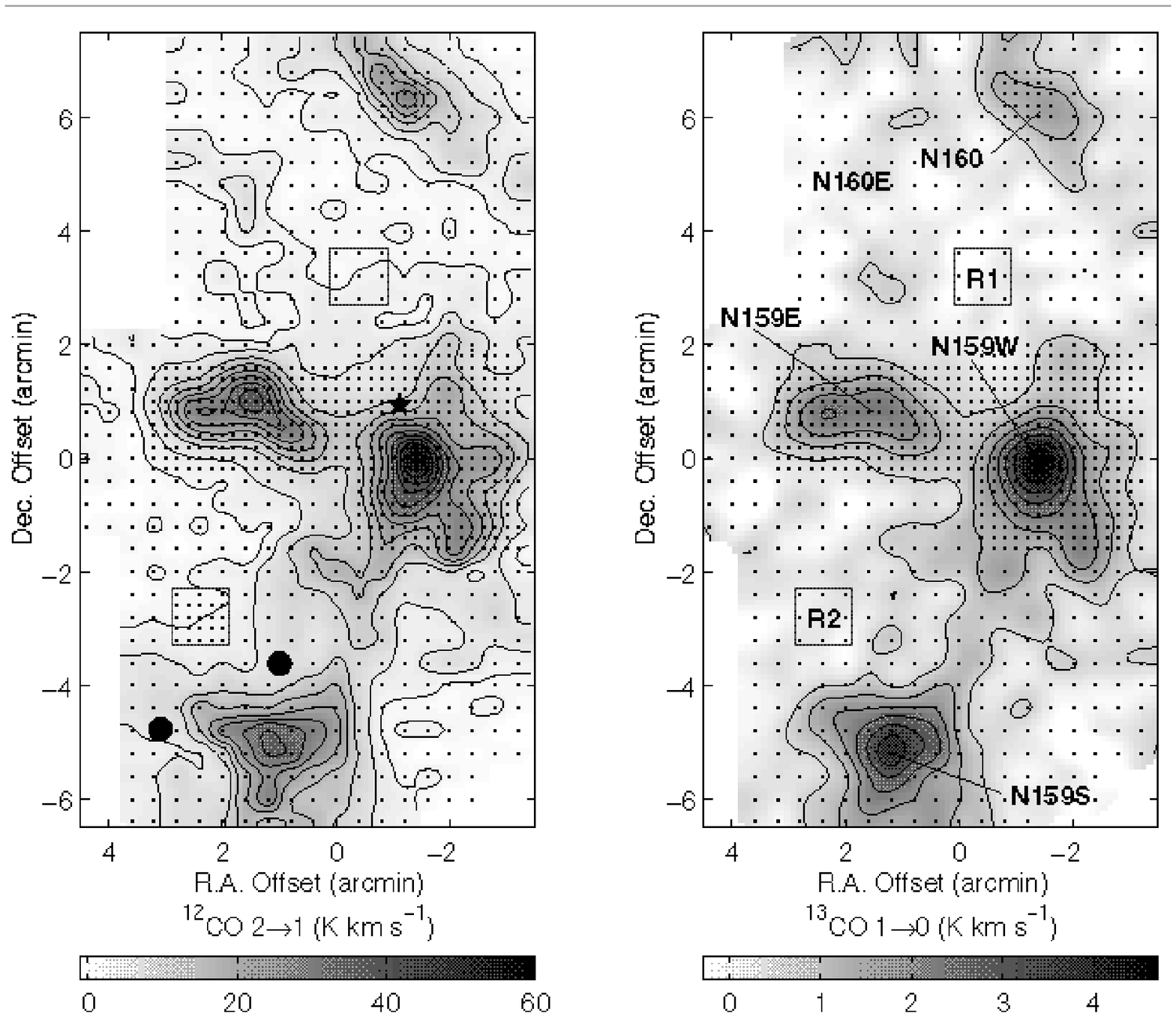} \figcaption[fig1.eps]{Carbon monoxide main beam integrated
intensity maps acquired at SEST. The \co\ \jtwo\ contours (left panel)
are 2, 5, 10, 15, \ldots, 55 \Kkmpers\ ($\sigma_{rms}\sim0.6$
\Kkmpers) and the data have been convolved with a Gaussian profile to
33\arcsec\ resolution. The \cothree\ \jone\ contours (right panel) are
0.6 to 4.4 at 0.6 \Kkmpers intervals ($\sigma_{rms}\sim0.2$ \Kkmpers)
and the map has been convolved to 60\arcsec\ resolution. The
individual pointing locations are shown by small dots. The boxes show
the position of the two regions with low level extended emission also
observed in \co\ \jone.  The map intensities have been integrated over
the velocity range 225 to 245 \kmpers. The black star denotes the
position of the black-hole candidate and strong X-ray source
LMC-X1. The black circles mark the positions of DEM L272 and L279, the
two \hii\ regions closest to N159S. The \cothree\ \jone\ map has been
annotated with the nomenclature used in Table \protect\ref{tabpos}.
\label{sestmaps}}
\end{figure}

\begin{figure}
\plotone{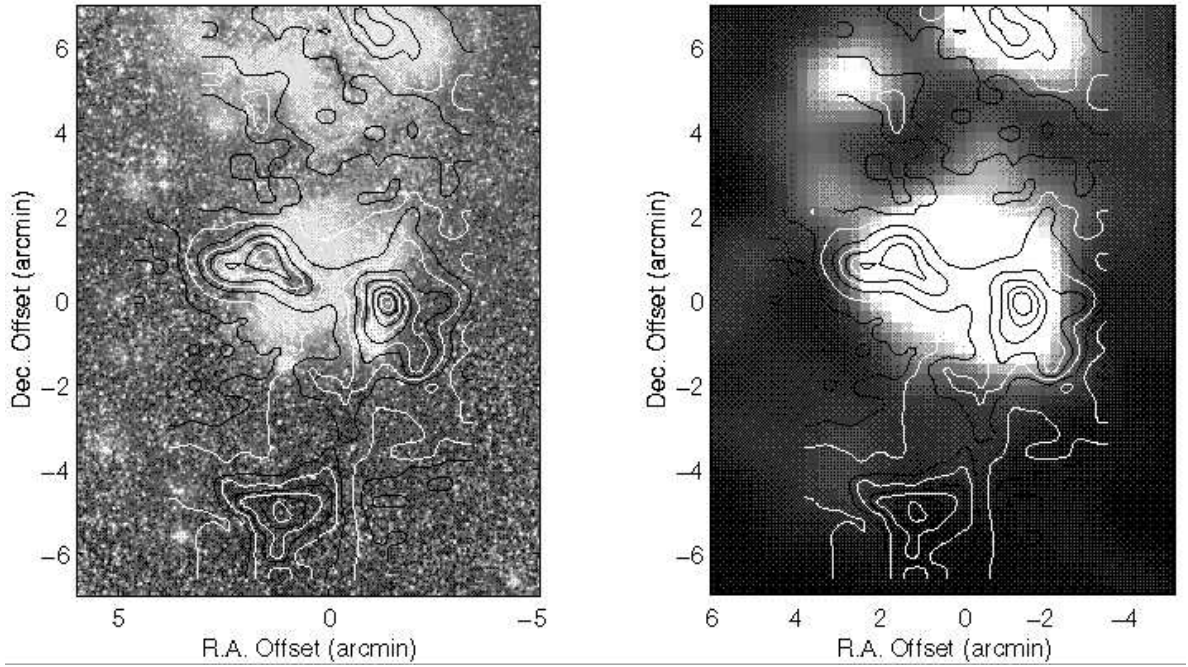} 
\figcaption[fig3.eps]{Heating of the molecular
cloud complex.  The optical (left panel, blue band, SERC) and 60
\micron\ FIR (right panel, IRAS HIRES, 40 iterations) emission are
shown here with \co\ \jtwo\ contours overlaid (alternate black and
white, same levels as Fig. \protect\ref{sestmaps}). This picture
illustrates the detailed structure of the nebulae and the intimate and
complex association between nebulosity, obscuration lanes, and
molecular clumps. The southern cloud appears to have no UV or heating
sources associated with it.\label{heating}}
\end{figure}

\begin{figure}
\plotone{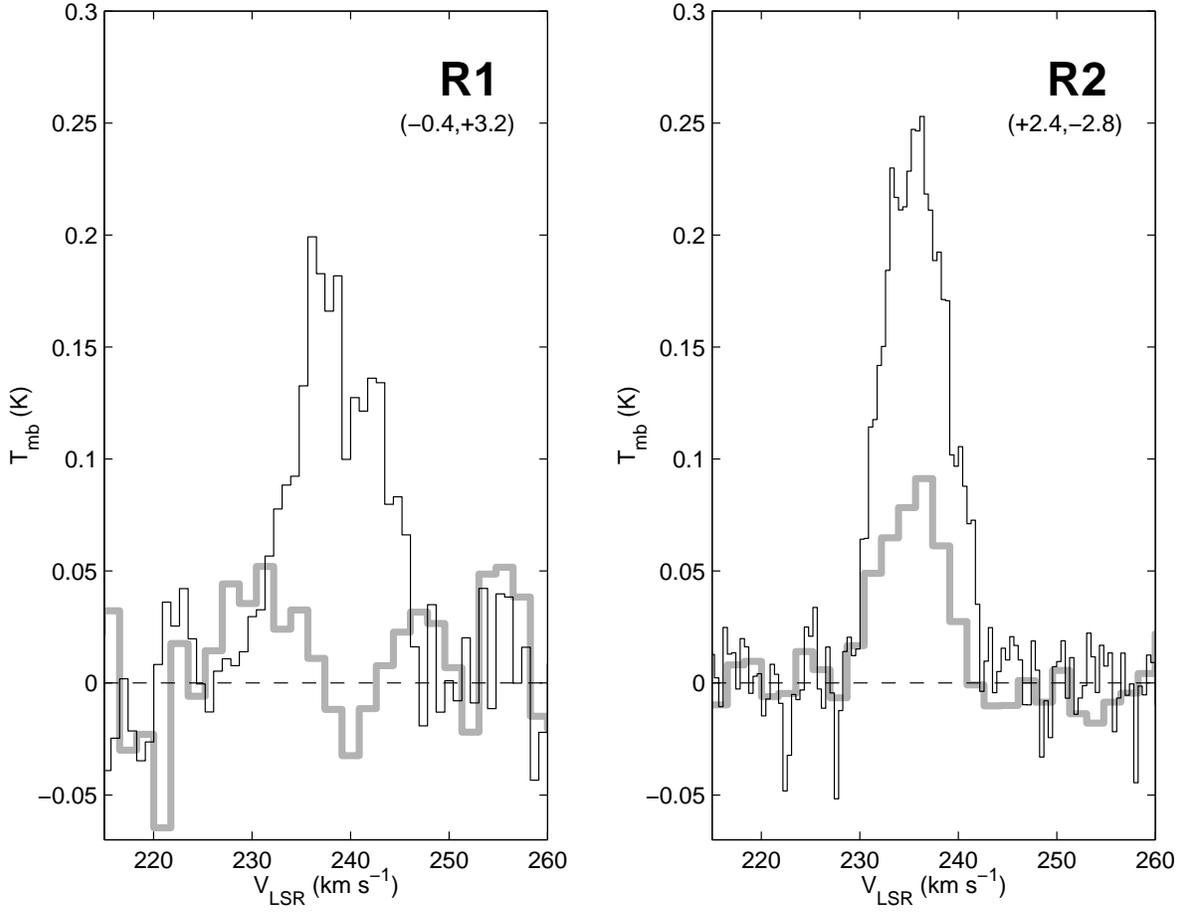} 
\figcaption[fig5.eps]{Observed spectra of the \co \jone\ and \jtwo\
transitions for the faint extended molecular emission. These two
regions are identified here by their offsets and drawn on
Fig. \protect\ref{sestmaps}. The \jtwo\ transition (fine black line)
produces the strongest lines in both regions while the \jone\ (wide
gray line) is weaker by a factor $\gtrsim3$, suggesting optically thin
emission.\label{cospectra}}
\end{figure}

To investigate the faint extended emission apparent in the \co\ \jtwo\
map, two small subregions at the edges of the complex (squares in
Fig. \ref{sestmaps}) were mapped simultaneously in \co\ \jtwo\ and
\jone. This permitted us to obtain reliable line ratios for these two
transitions. The average spectra for these two regions are shown in
Fig. \ref{cospectra}. The \co\ \jtwo/\jone\ ratio is $\sim3$ for both
positions; about a factor of 2---4 larger than for the molecular peaks
(cf., Table \ref{tabmm}). For thermalized, optically thick CO, we
expect a line ratio close to one. Larger \jtwo/\jone\ ratios can be
produced by essentially only four mechanisms: 1) self or
foreground-absorbed \jone\ emission, 2) optically thick CO gas with
temperature gradients, 3) different beam filling fraction for the
\jone\ and \jtwo\ transitions, or 4) thermalized but optically thin CO
emission.

\begin{figure}
\plotone{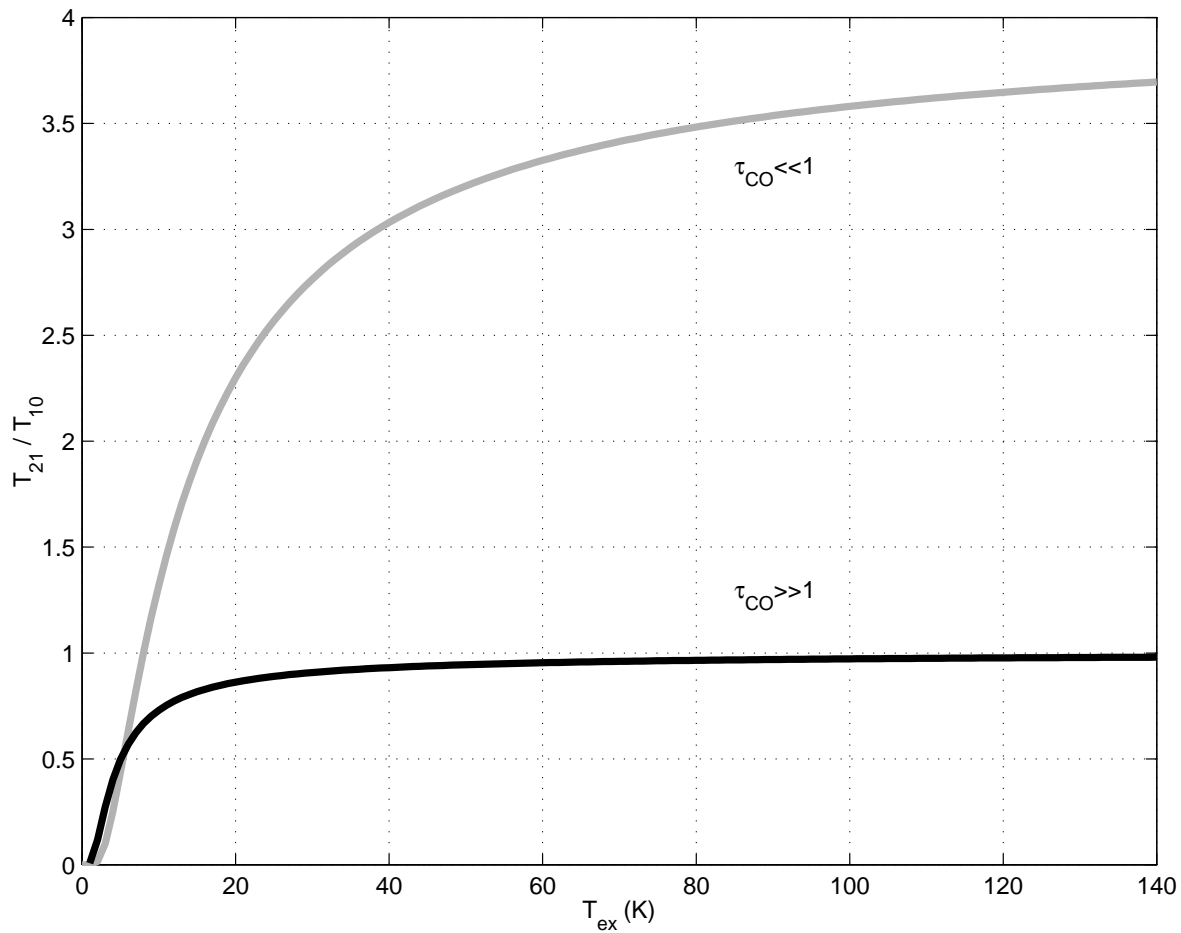}
\figcaption[fig6.eps]{Ratio of \jtwo\ to \jone\ integrated brightness 
temperature for
CO emission in local thermodynamic equilibrium (LTE), as a 
function of the excitation temperature $T_{ex}$. The black line corresponds
to optically thick emission. The grey line is the plot of Eq.
\protect\ref{thinratgen} for $n=1$, valid under optically thin conditions.
\label{linerat}}
\end{figure}

In the first of these cases the lower transition is absorbed by either
a cold cloud along the line of sight at similar velocities, or colder
outer layers of the same cloud.  If most of the CO in the absorbing
gas is in the ground state (i.e., the intervening cloud is very cold)
the \jtwo\ transition will not suffer this effect, thereby
artificially raising the \jtwo/\jone\ integrated intensity ratio. In
case 2) the $\tau=1$ surface arises at different places along the line
of sight for the two transitions. The observed line ratio will be the
ratio of temperatures of the regions where the $\tau=1$ surface
occurs. If there are temperature gradients in the cloud, then this
ratio can in principle take any value.  In case 3), because the
opacity grows faster with $N_{CO}$ in the \jtwo\ than in the \jone\
transition (Eq. \ref{rattau}), it is possible for the clumps to appear
larger in the higher transition and consequently fill more of the
beam. For this effect to be considerable it requires small and warm CO
clumps (cf., Appendix \ref{beamfill}).  In case 4) the ratio of both
transitions will be in the range 0---4, depending on the temperature
of the emitting CO gas (cf., Fig. \ref{linerat}).  It can be shown
that, in general, the ratio of integrated brightness temperatures for
two consecutive rotational transitions in the optically thin,
thermalized limit is

\begin{equation}
\frac{T_{{(J+1)}\rightarrow J}}{T_{J\rightarrow (J-1)}}=
{\left(\frac{J+1}{J}\right)}^2\,e^{-\frac{h\nu_{10}(J+1)}{kT_{ex}}}
\label{thinratgen}
\end{equation}

\noindent where $J$ is the rotational quantum number, $T_{ex}$ is the
excitation temperature, and $h\nu_{10}/k\sim5.5$~K for \co.  For the
transitions considered, a line ratio of $\approx3$ implies excitation
temperatures $T_{ex}\gtrsim40$~K.

Which one of these possibilities is occurring in regions R1 and R2?
The \jone\ spectrum for R2 (Fig. \ref{cospectra}) shows no clear
evidence for self-absorption.  Also, self-absorption tends to be a
rather localized phenomenon: it is difficult to imagine it happening
for two completely unrelated regions like R1 and R2. Thus we think it
is an unlikely explanation. To decide among the remaining
possibilities it would be ideal to be able to estimate the CO column
density. The \cothree\ observations, however, are not sufficiently
sensitive to provide useful limits for the \co/\cothree\ ratio and
subsequently determine if the emission is optically thick. 

Because some of the possibilities we are discussing require warm gas,
it is important to consider the sources of heating in these two
regions.  Region R1 is located between the N160 and N159 nebulae, and
consequently its temperature can easily be $>40$~K. Region R2 is
found, however, in a quiescent region about 4\arcmin\ ($\sim60$ pc
projected distance) away from the center of the N159 nebula.  To raise
the gas temperature of region R2 to the $\sim40$~K needed to explain
our ratios with optically thin CO emission, it is necessary to have
heating sources.  Surface temperature calculations of the PDR show
that only a modest radiation field, \xuv, is required to produce that
temperature (\xuv$\sim10$, Kaufman et al. 1999).  Eastward of N159S
there are a few very faint and inconspicuous \hii\ regions (DEM 272,
277 and 279, see Fig. \ref{sestmaps}; Davies et al. 1976), and traces
of low level 60 $\mu$m emission (Fig. \ref{heating}). This suggests
that photons, possibly from these \hii\ regions or from their much
brighter northern cousins, find their way there to heat the dust and
elevate the temperature of the diffuse gas. Calculations in
\S\ref{cidiscussion} show that the N159 nebula is probably bright
enough to provide the \xuv$\sim10$ field needed to raise the
temperature of the gas to 40~K.  This radiation is not, however,
intense enough to heat the core of N159S and make it a strong 60
$\mu$m or CO \jfour\ emitter (Fig. \ref{co43map}). It is certainly
not enough to produce any measurable 158 $\mu$m \cii\ emission, a
transition that requires $T\sim92$~K to be excited (Fig. \ref{cmap}).

Given the fact that regions R1 and R2 are at the edges of the
molecular cloud complex, we find the case for optically thin CO
emission more compelling than the alternative explanations.  It is,
however, difficult to choose among possibilities 2), 3), and 4) with
the available data. Nevertheless, we can make some specific
predictions for the last two cases.  In case 3) (larger beam-filling
fraction for the \jtwo\ transition), simple geometric arguments show
that very small CO clumps are required (Appendix
\ref{beamfill}). These clumps are so small that they have only
$\tau\sim2$ in the \jone\ transition and fill less than 1\% of the
beam. Future measurements of the \jthree\ and \jfour\ transitions in
these regions may help distinguish between case 3) and optically thin
emission (case 4). Because the ratio of opacities between the \jthree\
and \jtwo\ transitions of CO is 2.25 for infinite temperature
(Eq. \ref{thinratgen}), compared with 4 for $\tau$\jtwo/$\tau$\jone,
the increase in beam filling fraction going from the \jtwo\ to the
\jthree\ transition will be only modest. Thus, assuming uniform
density clumps in LTE we expect peak intensities $T_{mb}\sim0.18$ and
0.30~K in the \co\ \jthree\ transition for regions R1 and R2
respectively, only 20\% brighter than the observed \jtwo. For clumps
with density increasing toward the center the difference between the
beam filling fraction for both transitions, and consequently the
increase in brightness temperature, will be even smaller.

Concerning case 4), optically thin CO, the observed \jtwo\ intensities
require CO column densities of $N_{CO}\sim2\times10^{15}$ cm$^{-2}$
assuming optically thin LTE emission at $T_{ex}=40$~K. A prediction of
this model is that \co\ \jthree\ and \jfour\ emission should be
readily observable with peak intensities in a $\sim1\arcmin$ beam of
$T_{mb}\sim0.25$ and 0.45~K for regions R1 and R2 respectively.
Because of the low angular resolution of our \co\ \jfour\ data,
however, we are not able to cleanly separate these regions from the
nearby peaks that dominate the emission and therefore cannot test
these models. Nevertheless, the fact that these regions are relatively
bright in \ci\ (\S\ref{citocoratio}), together with the results of
column density calculations (\S\ref{coldens}), strongly suggest that
they are part of a translucent, mostly photodissociated envelope.

\subsection{Carbon in the Gas Phase of the ISM}
\label{cidiscussion}

\begin{figure}
\plotone{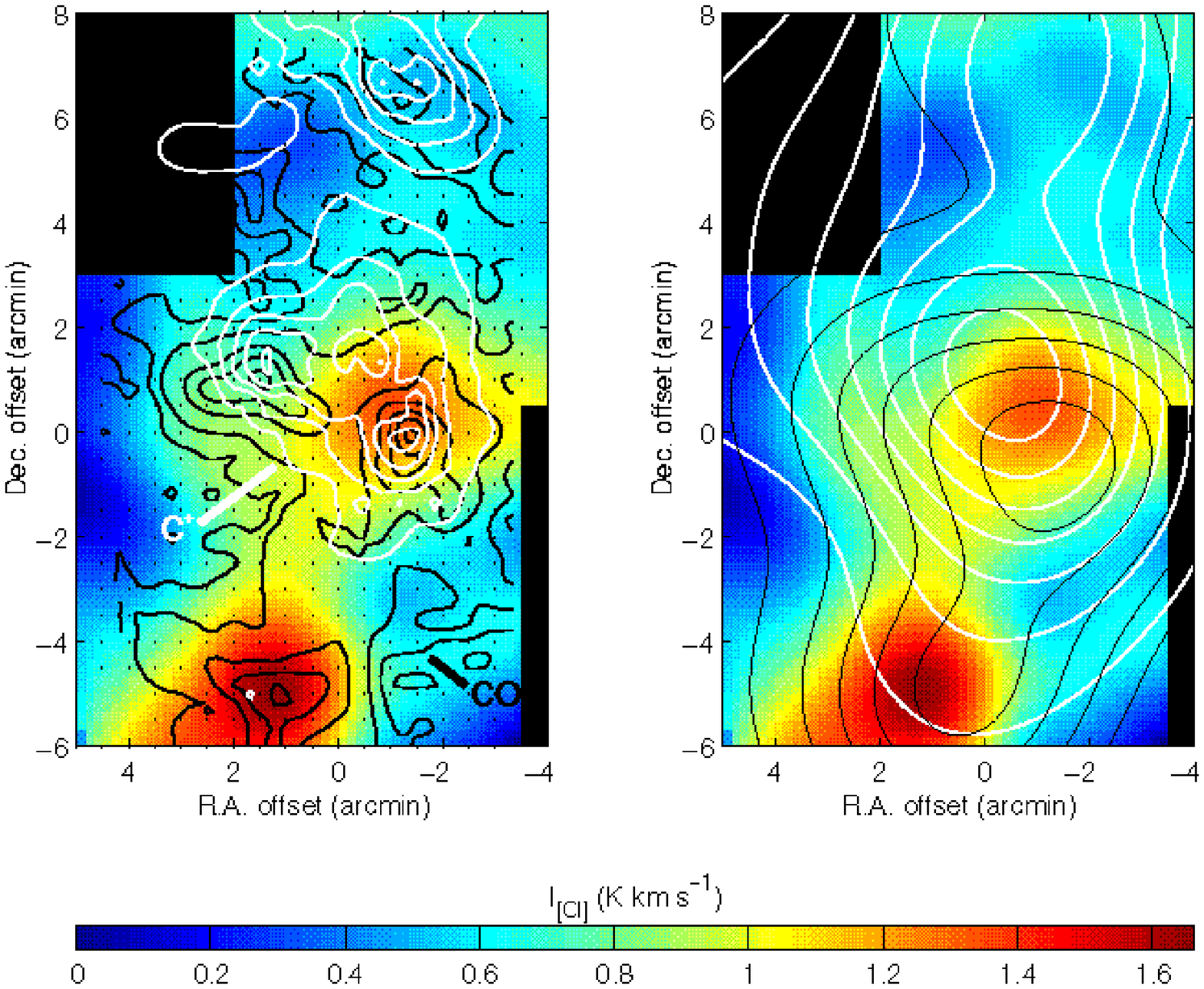} \figcaption[fig7.eps]{Distribution of the three
dominant forms of gas phase carbon in N159/N160. The pseudo color
image shows the AST/RO \ci\ observations at an angular resolution of
$\sim4.3\arcmin$ (sensitivity $1\sigma\sim0.1$ \Kkmpers). {\em (Left)}
In black contours is the CO \jtwo\ map, at a resolution
$\sim33\arcsec$ (contours 2,5,15,25,\ldots,55 \Kkmpers).  In white
contours is the \cii\ map by Israel et al.  (1996), at a resolution
$\sim55\arcsec$ (contours start at $6.8\times10^{-5}$ in steps of
$6.8\times10^{-5}$ \intunits). {\em (Right)} The three transitions
have been smoothed to a common resolution of 4.3\arcmin. We use here
the CO \jone\ map by Johansson et al. (1994) to avoid edge effects
when convolving the CO. The CO intensity contours start at 2 \Kkmpers\ 
in steps of 2 \Kkmpers. The \Icii\ contours start at $1\times10^{-5}$ 
in steps of $2\times10^{-5}$ \intunits.
\label{cmap}}
\end{figure}

Here we analyze the AST/RO \ci\ \fsone\ map in conjunction with
the available \cii\ and CO data (Israel et al. 1997; Johansson et al. 1998). 
Figure \ref{cmap} reveals the distribution of the three dominant forms of
carbon in the gas phase of the ISM. The striking features of this map are
the complex interplay between CO, \cn\ and \cp, and the fact that \ci\ peaks
in the southern cloud, a quiescent region entirely devoid of strong 
UV sources.

Throughout the map there is an overall anticorrelation between the
distributions of \cii\ and \ci.  Only in N159W do the three species
peak at approximately the same place, while in all the other clouds
only two of the species show intense emission.  The northern regions,
with abundant star formation activity, are bright in \cii\ but very
dim in \ci. The opposite is true for N159S.  This cloud, with no
massive star formation activity, is the peak of \ci\ in the whole
complex. It is also the region with the largest \ci/CO intensity ratio
(cf., Table \ref{tabsubmm}).  The \ci\ emission in N159E is relatively
faint, and the CO, \cn\ and \cp\ there appear to occupy adjacent and
partially overlapping regions.  Finally, in N160 a \ci\ hole is filled
in by a bright lobe of \cii, with a very faint ridge of \ci\ emission
overlapping with the CO.

\begin{figure}
\plotone{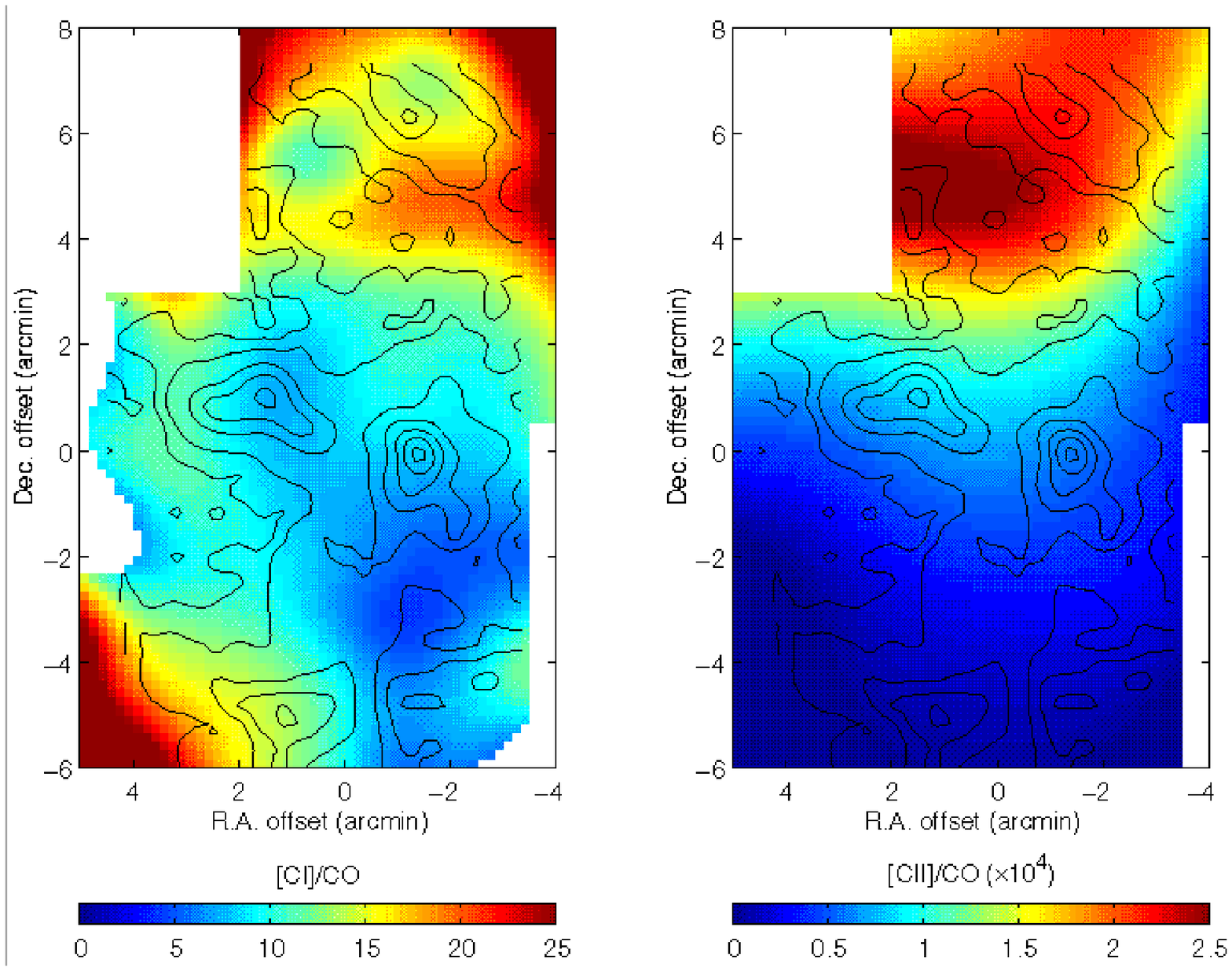} \figcaption[fig8.eps]{Maps of the \citoco\ and
\ciitoco\ ratios throughout the N159/N160 complex. {\em (Left)} Image
of the \citoco\ integrated intensity ratio, with \Ici\ and \Ico\
\jone\ in \intunits. The CO \jtwo\ contours are overlaid to serve as a
reference, but the \jone\ map from Johansson et al. (1994) was used to
produce the ratio, thereby eliminating edge problems in the CO
dataset.  {\em (Right)} Image of the \ciitoco\ integrated intensity
ratio. Typical Galactic ratios are \citoco$\sim10$ and
\ciitoco$\sim4400$. All maps have been convolved to the same
$\sim4.3$\arcmin\ angular resolution.\label{cratios}}
\end{figure}

The distribution of the different forms of gas phase carbon in
N159/N160 may be affected by peculiarities of the region. For example,
the central \cii\ peak (N159-M, Israel et al.  1996), which has no FIR
counterpart, is very close to the position of LMC-X1. It may be
associated with that X-ray source, which is located at offsets
$\sim[-1\arcmin,+1\arcmin]$ (Fig. \ref{sestmaps}). LMC-X1, one of the
strongest X-ray sources in the LMC, is a black-hole candidate with an
O7~III companion star (Cowley et al. 1995; Schmidtke, Ponder, \&
Cowley 1999).  It has been suggested that X-ray radiation has
considerable effect on the chemistry of the ISM (e.g., Lepp \&
Dalgarno 1996), and it is possibly a way to produce \cp\ and
\cn. Dissociation of CO in shocks may be another way of generating
these species. Approximately 2\arcmin\ east of LMC-X1, Chu et
al. (1997) have identified a supernova remnant, SNR 0540-697, based on
optical spectra and X-ray data. This remnant is expanding at
$\sim\pm150$ \kmpers\ and overlaps with most of the nebulosity NW of
N159E (Fig. \ref{heating}).  We see no evidence in our data for \ci\
emission at these velocities, and the ratio maps (Fig. \ref{cratios})
show no peculiar enhancement of the \citoco\ and \ciitoco\ ratios at
the position of the remnant. Thus, shock-induced dissociation of CO 
does not appear to be an important mechanism producing \cii\ or \ci\ 
in this area.

\begin{figure}
\plotfiddle{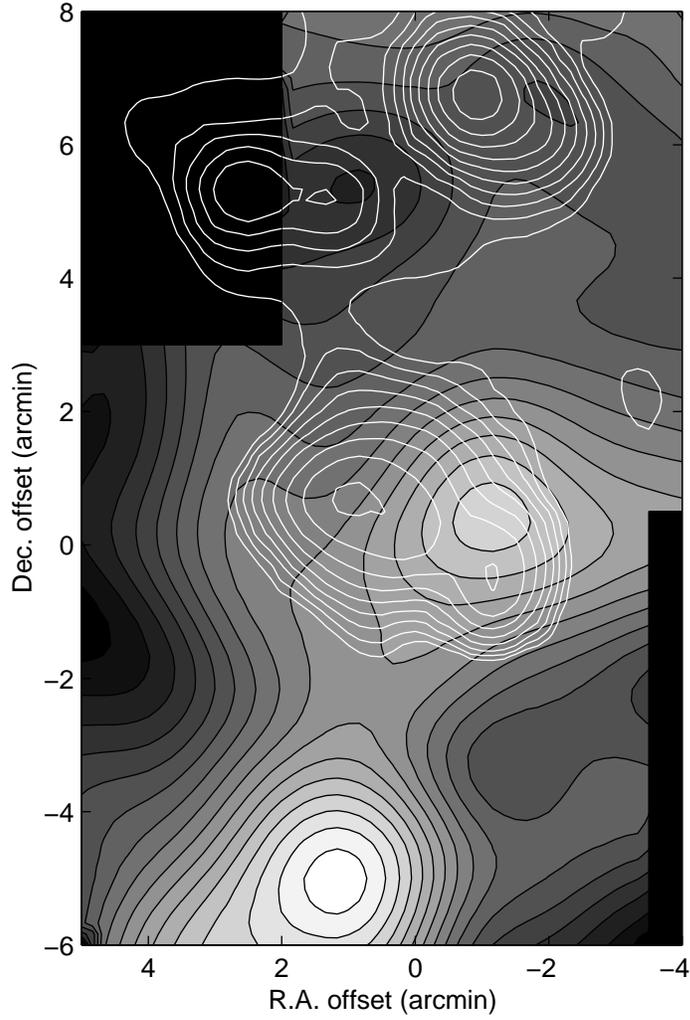}{4.5in}{0}{100}{100}{-315}{-200}
\figcaption[fig9.eps]{Maps of neutral carbon emission and 
21~cm radio continuum.
The radio continuum contours (white) start at 30 mJy and are logarithmically 
spaced by a factor of 1.4. The highest contour is at 620 mJy. The \Ici\ 
contours (grayscale) start at 0 \Kkmpers\ and go to 1.6 in steps 
of 0.1 \Kkmpers.\label{radio}} 
\end{figure}

On the spatial scale of these observations ($1\arcmin\sim15$ pc) we do
not expect to be resolving the PDRs into their three separate \cp,
\cn\ and CO regions. This has been accomplished only in a few Galactic
sources observable with much greater spatial resolution (e.g., the
Orion bar). Accordingly, we expect to observe coextensive \cii, \ci\ and
CO emission for the molecular peaks. This is not necessarily the case
for the translucent medium. In diffuse gas and in the presence of
strong UV sources most carbon will be ionized.  Thus \cp\ will
dominate the emission, forming \cii\ regions akin to Str\"omgren
spheres around the ionizing sources. This is apparently happening in
the northern portion of the map where bright lobes of \cii\ emission
east and west of N160 fill in holes in the \ci\ distribution. These
\cii\ peaks and \ci\ holes are unequivocally associated with \hii\
regions (Fig. \ref{radio}). Most of their \cii\ emission may 
arise from \cp\ mixed with \hii\ inside the Str\"omgren sphere, 
collisionally excited by electrons.

In principle, over a small range of extinction (\av$\sim0.3$---1) we
could have clouds where \cp\ and \cn\ are dominant, with little or no
CO emission. That may be occurring between N160 and N159W, where
bright \cii\ and \ci\ overlap in a region that shows little CO
emission. One must be careful, however, when interpreting and
combining data with very different (and, for the \ci, low) angular
resolutions.  Figure \ref{cmap}b shows the three transitions convolved
to a common resolution ($\sim4.3$\arcmin). In this map there are only
two peaks in the \cii\ distribution (N160 and N159W), and the \cii,
\ci\ and CO maxima near N159W are displaced by about 1\arcmin\ from
each other in a way that resembles a PDR \cp/\cn/CO transition. This
structure, however, is not a PDR: if these clouds were moved to Orion
(0.5 kpc away and perhaps the best example of a resolved PDR) the
distance between the \cii\ and the CO peaks would span 2 degrees in
the sky.  What we are observing are large scale excitation and
chemical gradients.  The extended \cii\ emission is heavily weighted
in the convolved data and pulls the overall maximum northward, also
diluting the peak coincident with N159W. In much the same manner, the
\jone\ peak is pulled southward by the extended CO emission. Notice
that the CO maximum near N159S moves westward for the same reason.

\subsubsection{The Neutral Carbon Emission from N159S}

The fact that the peak of \ci\ emission for the entire complex is a
quiescent region is unexpected if \cn\ has, chiefly, a PDR origin and
thus requires UV photons to be produced.  To review the evidence:
N159S is a dark cloud with weak CO \jfour\ emission (hence at a low
temperature), with no conspicuous heating sources apparent in the
optical or the FIR (hence little or no star formation), and with very
faint \cii\ emission (hence no UV sources). Nevertheless, according to
the analysis by Heikkil\"a et al. (1999) it is the cloud with the
largest column density in the complex ($N_{H_2}\sim1.7\times10^{22}$
cm$^{-2}$, \av$\sim4.5$ compared to $N_{H_2}\sim1.1\times10^{22}$
cm$^{-2}$, \av$\sim2.9$ for N159W and N160), and it is also the
brightest \ci\ emitter, possessing the largest \citoco\ ratio. This is
certainly indirect evidence, but it suggests that most of the \cn\ in
this cloud originates not by UV photodissociation of CO in the PDR,
but inside the CO cores.

Recent modeling results indicate, however, that the \citoco\ ratio is
very insensitive to the radiation field (Kaufman et
al. 1999). Therefore only a small amount of UV radiation is necessary
to explain the observed ratio in the context of a PDR. This can be
understood in the following way: in \av\ space the extent of the
\ci\ emitting region in a homogeneous one-dimensional calculation, and
consequently the \cn\ column density, is relatively insensitive to the
input radiation field. The depth at which the C$^+$/C$^0$/CO
transition occurs, however, increases for larger \xuv. Because \co\
becomes optically thick soon after this transition, neither the \ci\
nor the CO emerging intensities are a strong function of the radiation
field at the surface of the cloud. All the molecular peaks in this
region feature optically thick \co, based on the \cothree/\co\
intensity ratios listed in Table \ref{tabmm}
($^{12}$C/$^{13}$C$\sim50$ for the LMC, Johansson et al. 1994).

The previous reasoning applies to homogeneous clouds with
plane-parallel (i.e., one-dimensional) geometry.  If molecular clouds
are in fact clumpy, they can be modeled as an ensemble of spherical
cloudlets. In this scenario the UV field should have a dramatic effect
on the \cp, \cn\ and CO column densities and intensity ratios.  In a
clump subjected to an increasing UV field, the CO emitting region can
be pushed towards the center only as far as the radius of the clump,
after which the entire clump photodissociates. This scenario has been
modeled in detail by Bolatto et al. (1999). Following those
calculations we expect an inverse dependence between \xuv\ and CO
column density, and a modest increase in the \ci/CO ratio for
increasing \xuv\ (growing by $\approx25\%$ per order of magnitude in
\xuv). This model, however, does not include an interclump
medium. Modeling of clumpy PDRs as a collection of high-extinction
(\av$\sim100$), dense regions immersed in a diffuse interclump gas
exposed to high \xuv\ (Meixner \& Tielens 1993, 1995) suggests that
most of the CO \jone\ and \ci\ emission comes from the low density
interclump medium. Thus, at low UV fields and low clump filling
fractions, and when considering interclump dominated lines, a clumpy
PDR may behave very similarly to a homogeneous plane-parallel model of
interclump density. Similarly, at high clump filling fractions the
contribution from the interclump medium would be unimportant.

We will now try to estimate the radiation field in the N159S region.
This cloud is most likely illuminated by the bright northern \hii\
regions, although there may be a small contribution from the much
smaller nearby regions mentioned in \S\ref{translucentCO} 
(Figs. \ref{sestmaps}, \ref{heating}).  Estimates
for the luminosities and possible exciting sources of the FIR peaks in
the N159/N160 complex, based on IRAS HIRES data, are given in Table
\ref{tabfir}. These estimates are in reasonably good
agreement with the KAO results by Israel et al. (1996).  An equivalent
spectral type and multiplicity is assigned to each flux, by assuming
that all of the starlight is trapped by the dust and reradiated in the
FIR.  

In order to use these results to estimate the UV radiation field
far away from these sources we will assume that the radiation
intercepted by dust is only a fraction $f$ of the total luminosity,
and therefore a fraction $(1-f)$ escapes the star-forming region and
the enshrouding dust.  Furthermore, we will assume that about 50\% of
the luminosity of the exciting stars is emitted in the range
912---2066\AA\ (typical for late O/early B-type stars).  Therefore the
UV radiation field at a distance $d$ is

\begin{equation}
\xuv\approx\frac{(1-f) L_{FIR}}{8\pi f \chi_0 \, d^2}
\end{equation}

\noindent where $\chi_0$ is the standard normalization factor
(approximately the UV radiation field in the vicinity of the Sun,
$\chi_0=1.6\times10^{-3}$ erg s$^{-1}$ cm$^{-2}$; Habing 1968).  Using
this formula we can compute \xuv\ for N159S, assuming that it is
illuminated by the nearby N159 \hii\ region. At the projected distance
$d\sim90$ pc, and assuming that 90\% of the UV escapes the surrounding
dust, we obtain $\xuv\simeq60$. If only 50\% of the UV escapes then
the radiation field would be $\xuv\simeq6$.  By comparison, radiation
field estimates by Israel et al.  (1996) for the northern clouds
N159E-W and N160 near \hii\ regions range from 300---600 (cf., Table
\ref{tabpos}).  Our \xuv\ estimate for N159S is probably too large,
since: 1) we use the projected distance, and 2) we assume no dust
extinction between the northern \hii\ regions and N159S.

Another way of computing the luminosity of the UV sources in this
region is to use the radio continuum information.  The 21 cm radio
continuum data (Figure \ref{radio}) poses constraints similar to the
FIR on the luminosity of the stars that excite the \hii\ regions. In
Table \ref{tabradio} we have calculated the spectral type and
multiplicity, assuming optically thin free-free emission and following
Jackson \& Kraemer (1999) analysis.  These estimates are generally
lower than those based on the FIR, mostly because spectral types later
than O6 are still very luminous (adding to the FIR luminosity) but
contribute little to the radio continuum. Recent ISOCAM observations
(Comer\'on \& Claes 1998), for example, reveal three strong 15 $\mu$m
peaks in N159 which feature very faint radio continuum, and are thus
attributed to ultra-compact \hii\ regions which are optically thick at
21 cm.  These are examples of sources that would add to the FIR
luminosity but be invisible at 1.42~GHz.  Nevertheless, both the radio
continuum and the FIR measurements agree within a factor of $\sim2$
($L_{FIR}\sim2\,L_{RC}$), strongly suggesting that most UV is actually
intercepted by the surrounding dust. Taking into account the various
estimates and caveats discussed in the previous paragraph, we conclude
that \xuv(N159S)$\lesssim10$.  Such a low UV field is consistent with
the faint \cii\ emission from N159S, as well as its low \ciitoco\
ratio.

Even orders of magnitude in \xuv, however, have little impact on the
value of the \citoco\ ratio in standard PDR plane-parallel models.
According to the calculations by Kaufman et al. (1999), the \ci/CO
intensity ratio observed in N159S can be attributed to gas with
$n\approx 10^5$ \percmcu\ at \xuv$\sim$1---$10^5$. Note that
metallicity has only very small effects on plane-parallel model
calculations, for reasons essentially similar to those discussed for
\xuv. Thus, although the aforementioned calculations were performed
for Galactic sources their results can be safely applied to the LMC.
The coincidence between this density and that obtained by Heikkil\"a
et al. (1999) for N159S appears satisfying, until we take into account
the different scale of the measurements: the 4\arcmin\ (60 pc) region
over which \citoco$\simeq0.15$ is $\approx25$ times larger than the
SEST $\sim48\arcsec$ beam used to carry out the multiline excitation
analysis.  This is an extremely high density for such a large region.
Invoking beam filling fraction arguments avoids this problem, by
assuming that only small and dense regions within the beam are
dominating the CO \jone\ and \ci\ emission. Comparison of the measured
\ci\ intensity with the model results suggests $\sim10\%$ beam filling
fraction. Nevertheless, if such a high density is characteristic of the gas
producing the \ci\ emission, it rules out the interclump medium as the
main reservoir of neutral carbon. Meixner \& Tielens' (1995) modeling
of two-phase (clump + interclump) PDRs shows that \citoco$\sim10$ can
be obtained for a mixture of $5\times10^5$ \percmcu\ \av$\sim100$
clumps immersed in a $3\times10^3$ \percmcu\ diffuse component, with
20\% volume filling fraction for the clumps. These computations,
however, were carried out for \xuv$\sim5\times10^4$ and no similar
calculations are available for UV fields closer to that of N159S.

At the beginning of this section we pointed out a series of intriguing
facts about N159S that may suggest a non-PDR origin for most of its
neutral carbon. Despite them, however, we conclude that the observed
\cii, \ci, and CO intensities in N159S are consistent with standard
PDR theory assuming $n\approx10^5$ \percmcu\ and \xuv$\sim10$. These
values of the density and UV field agree with previous excitation
analysis and with the best estimates of \xuv\ in N159S, assuming it is
illuminated by the northern complex of \hii\ regions.

\subsubsection{The I$_{[CI]}$/I$_{CO}$ Ratio in the Complex}
\label{citocoratio}

How are the relative intensities of the different forms of carbon
affected by the local conditions? In the previous sections we
have seen that \ci\ is dim in the northern portion of the map,
where there is active emassive star formation, and bright in N159S.
In this section we will compare the \citoco\ and \ciitoco\ ratios 
throughout this region, and look for the effects of radiation
fields and metallicity.

Figure \ref{cratios} shows the \citoco\ and \ciitoco\ ratios mapped
for the entire complex, with the data convolved to a common
4.3\arcmin\ resolution.  Table \ref{tabsubmm} shows our \citoco\
results tabulated for the regions of interest. Notice that Table 
\ref{tabsubmm} uses a
4\arcmin\ beam size. To convert from intensity ratios given in
\Kkmpers\ to ratios in \intunits\ multiply by $\sim 78$ for
\citoco. Since N160 is such a weak \ci\ source, its ratio is 
dominated by the surrounding extended \ci\ emission and therefore the ratio
increases noticeably when measured in a larger beam. For N159S the
angular resolution of the measurement is unimportant, and the ratio
remains mostly unchanged. Therefore the high \citoco\ ratio at N159S
is not an artifact caused by the nearby emission.  The first
measurement of the \ci/CO intensity ratio in N159 was carried out by
Stark et al. (1997a), who obtained \citoco$\simeq0.26$ (intensities in
\Kkmpers). This determination was based on a single spectrum taken
towards the nominal position of N159W, and assumed an unresolved
structure for the CO from this cloud.  We find that, albeit in a
larger beam, \citoco$\simeq0.12$ for N159W. This difference is largely
due to an underestimate of the CO intensity in the aforementioned
paper.

Two regions of the map feature high \ci/CO intensity ratios: the
northern region, where there \ciitoco\ also peaks, and the southeast
corner, where there is extended low-level CO \jtwo\ but little \jone\
emission and no detectable \cii\ emission. We think that in both cases we are
seeing translucent, mostly photodissociated gas, with the important
difference that the abundant UV radiation in the northern region
raises the temperature of the PDRs over the 91.3~K required to excite
the 158 $\mu$m \cii\ transition (see \S\ref{coldens}).  There is the
possibility that the southeastern extension of the N159S cloud in \ci\
is an artifact of the sampling, since the map is missing a few spectra
there. Because of the heavy oversampling of the beam, however, we
think that this extension is probably real. It appears also that there
is a modest increase in the \citoco\ ratio at the edges of the
complex, and perhaps in the immediate vicinity of LMC-X1. The lowest
ratios are found south of N159W.

How does the \citoco\ ratio in this complex compare with the \ciitoco\
ratio?  Since both \cp\ and \cn\ are produced by the action of UV
photons in the PDR, the naive expectation is that \cii\ and \ci\
should have a similar distribution in unresolved PDRs.  Excitation
differences, important for very low radiation fields (\xuv$<10$), will
become negligible for \xuv$>100$ as the PDR reaches temperatures well
over 100~K (Kaufman et al. 1999). As discussed at the beginning of
\S\ref{cidiscussion} there are other reasons why we may not expect
\cii\ and \ci\ to be coextensive in the diffuse translucent gas,
namely the formation of \cp\ ``Str\"omgren spheres'' near UV
sources. In these regions carbon ionization and not dust absorption is
the dominant process that removes UV photons, and consequently all the
carbon is in the form of \cp.  In the molecular material, however, UV
photons should be predominantly removed by dust grains, leaving a
fraction of the carbon in the form of \cn. Consequently, if the
structure of the ISM is clumpy and therefore allows the UV photons to
penetrate deep into the clouds, the expectation is for \ci\ and \cii\
to be coextensive and in many ways behave similarly.  In particular,
modeling of clumpy PDRs suggest that both the \ciitoco\ and \citoco\
ratio should be enhanced in low metallicity systems (Bolatto et
al. 1999). This is mostly due to the rapidly dwindling sizes of the CO
cores of the clumps with decreasing metallicity and dust-to-gas ratio.

The \cii/CO intensity ratio has been studied in a variety of systems
(e.g., Stacey et al. 1991), and the highest ratios are associated with
low metallicity environments. As we pointed out in \S\ref{intron159},
Mochizuki et al. (1994) found a considerable enhancement of the global
\ciitoco\ ratio of the LMC (\ciitoco$\approx23,000$) over that of the
Milky Way. Average ratios for Galactic objects are \ciitoco$\sim1,300$
for GMCs, and $\sim4,400$ for \hii\ regions (Stacey et al. 1991).
Observations by Israel et al. (1996) of N159/N160 (cf., Table
\ref{tabpos}) show that the three northern molecular peaks (N159E,
N159W and N160) have \ciitoco\ ratios 1---5 times larger than
comparable Galactic regions while their I$_{\rm FIR}$/\Ico\ ratios are
3---40 times lower, suggesting a much larger abundance of \cp.
Measurements of the 30 Doradus region (Poglitsch et al. 1995) yielded
a ratio \ciitoco$\approx65,000$ for its molecular peaks.  Madden et
al. (1997) studied the global \cii\ emission from the low metallicity
dwarf galaxy IC~10 which has $Z_{\rm IC10}\approx Z_{\rm MW}/4$ (where we have
used the oxygen abundance in Orion as representative of the Galaxy,
$12+\log[{\rm O/H}]\approx8.75$; Lequeux et al. 1979). They found ratios in
the range \ciitoco$\sim$14,000---87,000 for various regions.

Unlike the widely varying \ciitoco\ ratio, the \citoco\ ratio for the
molecular peaks of N159/N160 is surprisingly uniform. The values range
only between \citoco$\sim7$ to $\sim12$. These are very similar to the
typical ratio in the Milky Way where \citoco$\sim13$ (intensities in
\intunits). This is not an isolated result. Wilson (1997) observed the
\ci\ emission from several clouds at different galactocentric
distances in M~33, a spiral galaxy with a well-known metallicity
gradient. She obtained ratios of \citoco$\sim3$---14 and found no
obvious trend with inferred metallicity.  Recently, Bolatto et
al. (2000) studied the \citoco\ ratio in the molecular cloud complex
associated with the brightest star-forming region of IC~10. They
obtained an average ratio \citoco$\sim18$, only slightly larger than
the average Galactic ratio. Finally, Gerin \& Phillips' (2000) recent 
study of atomic carbon in a variety of galaxies finds some dispersion in the
\citoco\ intensity ratio, but no clear segregation with galaxy type.
Their sample of galaxies has an average ratio \citoco$\sim16$, 
with most ratios found the in the interval $\citoco\sim$8---32.

Overall, if there is a trend for \citoco\ with metallicity it is
certainly much weaker than the one observed for \ciitoco.  A plausible
explanation for the constancy of the \ci/CO intensity ratio is a
non-PDR origin for a fraction of the \cn. While it seems unequivocal
that in the diffuse gas most neutral carbon is associated with PDRs,
it is possible that a different mechanism dominates its production
inside molecular cloud cores. If \cn\ is indeed produced in these
cores by processes unrelated to photochemistry, it would simply
explain the close observed association between the \ci\ and CO line
intensities. 

\subsection{The Column Density in the Extended Envelope}
\label{coldens}

The entire molecular cloud complex appears to be surrounded by an
extended envelope visible in CO \jtwo, and possibly \ci\ and \cii.
For example, the extended and relatively bright \ci\ emission east of
N159S and south of N159E correlates very well with the faint, extended
CO \jtwo\ in the same region, as does the tongue of \ci\ spreading
north of the N159 nebula. Assuming, as was discussed in
\S\ref{translucentCO}, that the CO emission in this region is
optically thin, then we appear to be seeing translucent gas emitting
in \ci. Translucent clouds are clouds with visual extinction
\av$\lesssim1$, where an important fraction of the carbon is in the
form of \cp\ and \cn\ (e.g., Ingalls et al. 1997). 
While the regions surrounding R1 are strong
\cii\ emitters, region R2 and its environs are very faint in
\cii. Undetected by Israel et al. (1996), the upper limit for the
integrated intensity from R2 is \Icii$\sim6.8\times10^{-5}$~\intunits.
The limit for the corresponding \cp\ column density, derived for the
high-temperature, high-density case (that is, the minimum column
density corresponding to the intensity limit) is
$N_{C^+}\lesssim4\times10^{16}$~\percmsq, about 20 times larger than
the corresponding column density of CO derived in
\S\ref{translucentCO}. Singly ionized carbon could well be the dominant
form of carbon in this region yet remain undetected.

Assuming that most of the carbon is in the form of \cp, we can
estimate the hydrogen column density along these lines of sight. For
R1, $\Icii\approx 1.4\times10^{-4}$ \intunits, therefore
$N_{C^+}\sim8\times10^{16}$~\percmsq. Assuming an LMC gas phase carbon
abundance per H nucleus $x_{C}\sim7\times10^{-5}$ (Dufour 1984), this
yields $N_{H}\sim1.2\times10^{21}$ \percmsq, or $\av\sim0.15$ (using
the conversion factor $1\,\av\sim8\times10^{21}$ \percmsq\ which is
four times that of the Galaxy). In the case of R2, where no \cii\ is
detected, Fig. \ref{cmap}b shows that there is some very low level
emission that becomes statistically significant only after spatial
smoothing.  We will assume \Icii$\sim3\times10^{-5}$ \intunits\ (about
half the sensitivity limit).  In the high-temperature, high-density
limit this would result in $N_{H}\sim4\times10^{20}$ \percmsq, or
$\av\sim0.05$. If we allow for a relatively low PDR temperature,
$T_{kin}\sim40$~K, consistent with a low \xuv\ and the optically thin
CO analysis, we find $\av\sim0.15$. It is important to point out that
these are line-of-sight extinctions, and not necessarily equal to the
\av\ that extincts the UV field and enters the PDR calculation. For a
completely edge-on PDR, for example, these two extinctions would be
measured along orthogonal directions and thus be unrelated.  If this
gas fills the beam and has a density near the critical density of
\cii\ ($n_{cr}\sim3\times10^3$ \percmcu), as we implicitly assumed in
the calculations, then we are seeing a $\sim0.1$ pc thick layer of
gas.  We find this sheet-like geometry, with a sheet thickness of only
1/100 of the extent in the plane of the sky, uncomfortable although
not impossible.

How do these results compare with column density predictions based on
the \hi\ data?  We assume a spin temperature of
$T_{sp}\sim110$---180~K based on the \hi\ absorption observed towards
the N159 continuum sources (Spitzer 1978). We can then compute atomic
hydrogen column densities of $N_{H}\sim4.2\times10^{21}$~\percmsq\ and
$N_{H}\sim4.5\times10^{21}$~\percmsq\ for regions R1 and R2
respectively.  These column densities are $\sim3$ times larger than
those derived from \Icii. The gas is apparently hot enough to populate
the upper fine structure level of \cp\ and excite the 158 \micron\
\cii\ transition (recall $h\nu/k=91.3$~K for \cii), thus this
discrepancy is probably not due to temperature effects.

This method, however, is not
very precise since it heavily relies on similar excitation conditions
for the \hi\ seen in absorption and emission, and this assumption is
compromised by the large scales involved in the map (recall
1\arcmin$\sim15$ pc).  Other possible causes for the apparent
discrepancy in $N_{H}$ are: 1) a deficiency of \cp\ in the atomic gas
(only 30\% of the C is photoionized), 2) the volume density of the
atomic gas is low, and consequently the excitation of \cii\ is
subthermal ($n<n_{cr}(\cii)$), or 3) there are density fluctuations
within the beam, and the \cii\ emission is dominated by the fraction
of the \cp\ column that is thermalized ($n>n_{cr}(\cii)$). 

The first possibility appears to be very improbable, since unless the
material is well shielded from the UV most of the carbon should be in
the form of \cp\ rather than \cn.  Unfortunately the large beam in our
\ci\ observations makes impossible to separate cleanly these regions
from the molecular peaks, thus precluding us from obtaining a reliable
estimate for the column density of \cn\ in R1 and R2. Subthermal
excitation of \cp\ requires the density of \hi\ to be below the
critical density of the 158 \micron\ transition, $n_{cr}\approx3000$
\percmcu. The intensity \Icii\ of \cp\ collisionally excited by H
atoms can be computed using (e.g., Madden et al. 1997):

\begin{equation}
\Icii=2.35\times10^{-21}\,N_{C^+}\left[\frac{2\exp(-h\nu/kT)}{1+2\exp(-h\nu/kT)+n_{cr}/n}\right]
\end{equation}

\noindent where $h\nu/k=91.3$ K. Consequently, a discrepancy of
a factor of 3 would require $n\sim500$ \percmcu, well below the
critical density of \cii. Notice that this density would bring the
physical thickness of the \cii\ emitting layer to $\sim3$ pc. The
third possibility (density fluctuations within the beam, akin to
clumping) would make the regions where $n>n_{cr}$ dominate the
\cii\ emission (about 1/3 of the total column), while in most of the
gas $n\ll n_{cr}$ and the \cii\ excitation is subthermal.  This ought
to be happening at some level, because CO \jtwo\ is present throughout
the region and its excitation requires $n\sim10^4$ \percmcu.  This
density is probably too large to be the average volume density in the
envelope, thus most of the CO and perhaps a large fraction of the
\cii\ emission is originating in clumps within the envelope. 

\section{Summary and Conclusions}

We have discussed new \cothree\ \jone, \co\ \jtwo, \co\ \jfour, and 
\ci\ \fsone\ emission line maps of the N159/N160 molecular cloud complex of
the Large Magellanic Cloud. 

The \co \jtwo\ map shows extended faint emission previously
undetected in the \jone\ transition. Further analysis of the
\jtwo/\jone\ intensity ratio in two selected subregions (R1 and R2)
shows that this ratio is $\gtrsim3$ for the extendend low level
emission.  This high value indicates optically thin CO emission from
warm gas ($T_{kin}\geq40$~K).  The \citoco\ ratio for R1 and R2
appears modestly enhanced with respect to the molecular peaks
(Fig. \ref{cratios}).  The \ciitoco\ ratio is large for the northern
region (R1), but \cii\ was not detected by Israel et al. (1996)
towards region R2. Because of the low UV field heating the ISM in the
southern region, this may be a temperature effect. The gas in the
southern portion of the complex, far away from the massive star formation and 
bright \hii\ regions, is too cold to excite the 158 \micron\ transition
($T<92$~K).  The faint CO envelope appears to be 
translucent ($\av<1$), and column density calculations assuming
that \cp\ is the dominant form of carbon confirm this result.  The
column density derived from the neutral hydrogen 21 cm data appears to
be $\sim3$ times larger than $N_{H}$ obtained from \cii. This is
probably caused by density enhancements (clumping) within the beam.

The \co\ \jfour\ map shows that the \jfour/\jone\ intensity
ratio is $\sim4$ times larger in the northernmost cloud (N160) than in
the southern cloud (N159S). This agrees very well with
indicators of star formation activity (I$_{\rm FIR}$, \Icii, radio
continuum), and is probably due to the much lower temperature
of the quiescent southern cloud. Estimates of the radiation field
incident on N159S suggest \xuv$\lesssim10$, consistent with a low
temperature. The ratio of 100/60 \micron\ continuum also indicates
a low temperature.

The \ci\ \fsone\ map shows the \ci\ intensity and the \citoco\
ratio for the molecular concentrations peaking in N159S.  The
observed values are consistent with the PDR computations available for
homogeneous, plane-parallel models, at $n\sim10^{5}$ \percmcu\ and
$\xuv\sim10$. Because in those models \Ici\ is not sensitive to \xuv,
however, the presence of intense \ci\ emission in N159S is not a good
test for a possible non-PDR origin for \cn\ inside GMCs. There is
pervasive \ci\ emission throughout the region between N159EW and
N160. This region also emits strongly in \cii, and features some of
the largest \citoco\ and \ciitoco\ ratios in the complex. The radio
continuum sources embedded in this gas appear as holes in the \ci\
distribution, while they are peaks in the \cii.  The \cii\ in these peaks
probably originates inside the \hii\ regions, mostly excited
by collisions with electrons.

The \ci/CO intensity ratios measured in the molecular peaks of the
complex range between $\citoco\sim7$---12 (intensities in
\intunits). This ratio is similar to the average ratio in the Milky
Way (\citoco$\approx13$) and to that measured in the low metallicity
dwarf IC~10 ($\citoco\sim18$, Bolatto et al. 2000).  The \ci/CO
intensity ratio for the translucent gas is somewhat larger,
$\citoco\sim20$---30. While intense \cii\ \fscii\ emission and large
\ciitoco\ ratios appear to be unequivocally associated with massive
star formation, \ci\ \fsone\ emission and the \citoco\ ratio shows no
such clear association and has a much more complex behavior. The
\citoco\ ratio is more uniform than the \ciitoco\ ratio throughout the
complex, where spans values of $\citoco\sim$5---25 whereas \ciitoco\
ranges between $\sim$500---25,000 (i.e., a factor of 5 versus a factor
of 50). This is partially due to the more stringent excitation
conditions required by \cii.

We believe that our understanding of the ISM, especially the actively
star-forming ISM, can be enormously advanced by further studies of the
Magellanic Clouds. These galaxies, with their proximity, their active
star formation, their low metallicities, and their unobscured lines of
sight, present unique opportunities for detailed multiwavelength
studies. Among the different regions in the Clouds, the N159/N160
complex is located at a privileged place, between the violent
starburst of 30~Doradus and the quiescent southern CO ridge. Because
of its location, its distinct environments, and the wealth of
phenomena taking place within its bounds, the N159/N160 complex is one
of the most interesting places in the Magellanic
Clouds. Unfortunately, these galaxies are out of the reach of the
large radioastronomy facilities in the Northern Hemisphere. The advent
of some of the observatories now being planned, such as the Atacama
Large Millimeter Array or a large South Pole telescope, will change
that. The study of the ISM in these objects will vastly benefit from
large submillimeter telescopes, array receivers, and
millimeter/submillimeter interferometers located south of the
equator. Future studies of N159/N160 should include the detailed
characterization of the translucent molecular envelope, the dust
properties, and the heating and cooling balance of the ISM near and
away from star formation sites. Other observational programs could
address the creation of molecular clouds by ram pressure compression,
and their disruption by star formation activity. 

\acknowledgments We wish to thank Thomas M. Bania for his thorough
review of the draft of this manuscript, the SEST crew at La
Silla for a smooth observing run, and the rest of the AST/RO team for
their support.  The research of A.D.B., J.M.J, and
X.Z. was supported in part by the National Science Foundation under a
cooperative agreement with the Center for Astrophysical Research in
Antarctica (CARA), grant number NSF OPP 89-20223.  CARA is a National
Science Foundation Science and Technology Center.  The research of
A.D.B. and J.M.J. was also supported in part by the National Science
Foundation through grant AST-9803065.  This research has made use of
NASA's Astrophysics Data System Bibliographic Services, the Los Alamos
National Laboratory preprint database, and the Centre de Donn\'ees
astronomiques de Strasbourg databases.

\newpage
\appendix
\section{Beam filling ratio for two transitions}
\label{beamfill}

For this calculation we will assume spherical clumps of radius $R$,
immersed in an isotropic UV field.  A transition turns optically thick
(i.e., its opacity, $\tau$, is unity) at a distance $x$ from the
surface of the clump. Therefore the radius of the clump at given
transition (i.e., the distance from the center of the clump to the
transition's $\tau=1$ surface) is $r=R-x$.  The ratio of beam filling
fractions $\Phi$ in two transition will then be proportional to the
ratio of the projected areas

\begin{equation}
\frac{\Phi_2}{\Phi_1}={\left(\frac{R-x_2}{R-x_1}\right)}^2
\label{phirat}
\end{equation}

It can be shown easily that for gas in LTE 
the ratio of optical depths in the \jone\ and \jtwo\ transitions will be

\begin{equation}
\frac{\tau_{21}}{\tau_{10}}=2\frac{\left({1-e^{-\frac{h\nu_{21}}{kT_{ex}}}}\right)}
{\left({e^{\frac{h\nu_{10}}{kT_{ex}}}-1}\right)}
\label{rattau}
\end{equation}

\noindent where $h$ and $k$ are Planck's and Boltzmann's constants
respectively, and $T_{ex}$ is the excitation temperature.  For uniform
density clumps the $\tau=1$ surface occurs first for the transition
with faster growing opacity, that is, $x_2/x_1=\tau_1/\tau_2$.
Equation \ref{phirat} thus has the solution

\begin{equation}
R=x_1\frac{\sqrt{\frac{\Phi_2}{\Phi_1}}-\frac{\tau_1}{\tau_2}}{\sqrt{\frac{\Phi_2}{\Phi_1}}-1}
\end{equation}

In order to reproduce the CO \jtwo/\jone\ intensity ratios observed in
regions R1 and R2, $\Phi_2/\Phi_1\simeq3$. Using
$\tau_2/\tau_1\simeq3$ (i.e., $T_{ex}\sim40$~K according to
Eq. \ref{rattau}) yields $R\sim1.9\, x_1$. Thus
$\tau(1\rightarrow0)\sim2$ at the center of the clump. The average
over the whole projected spherical clump is $\sim30$\% larger, or
$\tau(1\rightarrow0)\sim2.5$.

\newpage

\newpage

\newpage

\newpage

\newpage

\newpage

\newpage

\newpage

\newpage

\newpage
\begin{deluxetable}{lcccccccr}
\tablewidth{0pc}
\scriptsize
\tablecaption{Positions and Properties of the Regions of Interest in N159/N160}
\tablehead{
Identifier &    R.A. 	&   Dec.	& \multicolumn{1}{c}{R.A. offset} & 
\multicolumn{1}{c}{Dec. offset} & \multicolumn{1}{c}{$n($H$_2)$\ 
\tablenotemark{a}} & \multicolumn{1}{c}{N$_{\rm H_2}$\ \tablenotemark{a,d}} & 
\multicolumn{1}{c}{$\chi_{uv}$\ \tablenotemark{b,c}} & 
\multicolumn{1}{c}{\ciitoco\ \tablenotemark{b}}\\
&   (B 1950) 	&  (B 1950)  	& \multicolumn{1}{c}{(arcmin)} & 
\multicolumn{1}{c}{(arcmin)} & \multicolumn{1}{c}{(cm$^{-3}$)} &
\multicolumn{1}{c}{(cm$^{-2}$)} & \multicolumn{1}{c}{}\\
}
\startdata
N159W	& $5^{\rm h}40^{\rm m}02\fs2$	&$-69$\arcdeg47\arcmin06\arcsec & $-1.4$ & $-0.1$ & $3\times10^5$ & $1.1\times10^{22}$ & 600---250 & 5600\\
N159E	& $5^{\rm h}40^{\rm m}35\fs8$	&$-69$\arcdeg46\arcmin12\arcsec & $+1.5$ & $+0.8$ & \ldots & \ldots & \ldots & 13000\\
N159S	& $5^{\rm h}40^{\rm m}27\fs7$	&$-69$\arcdeg52\arcmin00\arcsec & $+0.8$ & $-5.0$ & $1\times10^5$ & $1.7\times10^{22}$ & 60---6 & 400\\
N160	& $5^{\rm h}40^{\rm m}06\fs9$	&$-69$\arcdeg40\arcmin30\arcsec & $-1.0$ & $+6.5$ & $5\times10^5$ & $1.1\times10^{22}$ & 450---200 & 34000\\
N160E   & $5^{\rm h}40^{\rm m}46\fs1$	&$-69$\arcdeg41\arcmin48\arcsec & $+2.8$ & $+5.2$ & \ldots & \ldots & \ldots & \multicolumn{1}{r}{\ldots}\\
R1   & $5^{\rm h}40^{\rm m}13\fs8$	&$-69$\arcdeg43\arcmin48\arcsec & $-0.4$ & $+3.2$ & \ldots & \multicolumn{1}{r}{$6\times10^{20}$} & 200---50 & \multicolumn{1}{r}{\ldots}\\
R2   & $5^{\rm h}40^{\rm m}46\fs2$	&$-69$\arcdeg49\arcmin48\arcsec & $+2.4$ & $-2.8$ & \ldots & \multicolumn{1}{r}{2---6$\times10^{20}$} & 100---10 & \multicolumn{1}{r}{\ldots}\\

\tablenotetext{a}{From the mm multiline excitation analysis by 
Heikkil\"a et al. (1999).}
\tablenotetext{b}{From Israel et al. (1996).}
\tablenotetext{c}{From this work.}
\tablenotetext{d}{To convert to visual extinction, \av, divide by $4\times10^{21}$.}
\enddata
\label{tabpos}
\end{deluxetable}

\newpage
\begin{deluxetable}{lr@{$\pm$}lr@{$\pm$}lr@{$\pm$}lr@{$\pm$}lr@{$\pm$}l}
\tablewidth{0pc}
\scriptsize
\tablecaption{Measured Millimeter Line Parameters in N159/N160}
\tablehead{
\multicolumn{1}{c}{Identifier} & \multicolumn{2}{c}{T$_{mb}$\ \tablenotemark{a,b}} & \multicolumn{2}{c}{V$_{\rm LSR}$\ \tablenotemark{b}} & \multicolumn{2}{c}{FWHM\ \tablenotemark{b}} & \multicolumn{2}{c}{I=$\int {\rm T}_{mb}\,dv$\ \tablenotemark{a}} & \multicolumn{2}{c}{I/I$_{10}$\ \tablenotemark{c,d}}\\
&  \multicolumn{2}{c}{(K)} & \multicolumn{2}{c}{(km s$^{-1}$)} & \multicolumn{2}{c}{(km s$^{-1}$)} & \multicolumn{2}{c}{(K km s$^{-1}$)} & \multicolumn{2}{c}{}\\
}
\startdata
\sidehead{$^{13}$CO ($J=1\rightarrow0$)\ \tablenotemark{e}}
N159W 	& 0.64 & 0.01 & 238.1 & 0.1 & 6.9 & 0.1 & 4.67 & 0.04 & 0.12 & 0.02 \\
N159E 	& 0.60 & 0.01 & 234.6 & 0.1 & 6.1 & 0.1 & 2.13 & 0.04 & 0.07 & 0.01 \\
N159S	& 0.57 & 0.01 & 235.6 & 0.1 & 5.7 & 0.1 & 3.37 & 0.07 & 0.10 & 0.01 \\
N160	& 0.56 & 0.01 & 237.7 & 0.1 & 4.6 & 0.2 & 1.39 & 0.06 & 0.12 & 0.02 \\
R1      & \multicolumn{2}{c}{$\ldots$} &\multicolumn{2}{c}{$\ldots$} &\multicolumn{2}{c}{$\ldots$} & 0.01 & 0.10 & \multicolumn{2}{c}{$\ldots$}\\
R2      & \multicolumn{2}{c}{$\ldots$} &\multicolumn{2}{c}{$\ldots$} &\multicolumn{2}{c}{$\ldots$} & 0.09 & 0.11 & \multicolumn{2}{c}{$\ldots$}\\
\sidehead{$^{12}$CO ($J=2\rightarrow1$)\ \tablenotemark{e}}
N159W 	& 4.80 & 0.01 & 237.7 & 0.1 & 7.8 & 0.1 & 39.12 & 0.06 & 1.01 & 0.14 \\
N159E 	& 3.60 & 0.01 & 234.6 & 0.1 & 7.3 & 0.1 & 27.42 & 0.06 & 0.89 & 0.13 \\
N159S	& 3.22 & 0.01 & 234.9 & 0.1 & 8.1 & 0.1 & 27.00 & 0.08 & 0.78 & 0.11 \\
N160	& 3.46 & 0.01 & 237.7 & 0.1 & 5.0 & 0.1 & 18.24 & 0.10 & 1.59 & 0.23 \\
R1      & 0.16 & 0.01 & 238.0 & 0.7 & 10.8 & 1.5 & 1.75 & 0.20 & 3.57 & 1.40 \\
R2      & 0.25 & 0.01 & 235.4 & 0.2 & 7.6  & 0.4 & 1.94 & 0.10 & 2.81 & 0.64 \\

\tablenotetext{a}{Errors are statistical and $\pm1\sigma$. 
Systematic calibration uncertainty is $1\sigma\sim\pm10$\%.}
\tablenotetext{b}{Values and errors derived from Gaussian fit.}
\tablenotetext{c}{Ratio of transition intensity to that of 
\co \jone, convolved to the same beam, in \Kkmpers.}
\tablenotetext{d}{Errors are $\pm1\sigma$, and include 
10\% $1\sigma$ calibration uncertainty in both lines.}
\tablenotetext{e}{Measured in a HPBW=1\arcmin\ beam.}
\enddata
\label{tabmm}
\end{deluxetable}

\newpage
\begin{deluxetable}{lr@{$\pm$}lr@{$\pm$}lr@{$\pm$}lr@{$\pm$}lr@{$\pm$}lr}
\tablewidth{0pc}
\scriptsize
\tablecaption{Measured Submillimeter Line Parameters in N159/N160}
\tablehead{
\multicolumn{1}{c}{Identifier} & \multicolumn{2}{c}{T$_{mb}$\ \tablenotemark{a,b}} & \multicolumn{2}{c}{V$_{\rm LSR}$\ \tablenotemark{b}} & \multicolumn{2}{c}{FWHM\ \tablenotemark{b}} & \multicolumn{2}{c}{I=$\int {\rm T}_{mb}\,dv$\ \tablenotemark{a}} & \multicolumn{2}{c}{I/I$_{10}$\ \tablenotemark{c,d}} & \multicolumn{1}{c}{\ciitoco\ \tablenotemark{e}}\\
&  \multicolumn{2}{c}{(K)} & \multicolumn{2}{c}{(km s$^{-1}$)} & \multicolumn{2}{c}{(km s$^{-1}$)} & \multicolumn{2}{c}{(K km s$^{-1}$)} & \multicolumn{2}{c}{}\\
}
\startdata
\sidehead{$^{12}$CO ($J=4\rightarrow3$)\ \tablenotemark{f}}
N159W 	& 0.95 & 0.01 & 237.7 & 0.1 & 8.6 & 0.1  & 8.43 & 0.08 & 0.58 & 0.09 \\
N159E 	& 0.42 & 0.01 & 235.6 & 0.1 & 10.2 & 0.2 & 4.42 & 0.09 & 0.48 & 0.07 \\
N159S	& 0.40 & 0.01 & 235.5 & 0.1 & 7.7 & 0.2  & 3.41 & 0.08 & 0.27 & 0.04 \\
N160	& 0.39 & 0.07 & 239.1 & 0.1 & 9.4 & 0.2  & 3.81 & 0.08 & 1.16 & 0.17 \\
\sidehead{C$^0$ ($^3$P$_1\rightarrow^3$P$_0$)\ \tablenotemark{f}}
N159W   & 0.13 & 0.01 & 237.6 & 0.1 & 11.8 & 0.4 & 1.59 & 0.05 & 0.11 & 0.02 & 6000\\
N159E   & 0.14 & 0.01 & 235.5 & 0.1 & 6.1 & 0.3  & 0.87 & 0.03 & 0.09 & 0.01 & 7900\\
N159S   & 0.17 & 0.01 & 238.1 & 0.1 & 10.1 & 0.4 & 1.90 & 0.06 & 0.15 & 0.02 & 1000\\
N160    & 0.14 & 0.01 & 239.2 & 0.1 & 2.4 & 0.1  & 0.34 & 0.02 & 0.10 & 0.02 & 21700\\
R1      & 0.06 & 0.01 & 238.3 & 0.3 & 9.1 & 0.7  & 0.55 & 0.04 & 0.14 & 0.02 & 17800\\
R2      & 0.09 & 0.01 & 236.6 & 0.2 & 7.7 & 0.6  & 0.73 & 0.05 & 0.12 & 0.02 & 1700\\

\tablenotetext{a}{Errors are statistical and $\pm1\sigma$. 
Systematic calibration uncertainty is $1\sigma\sim\pm10$\%.}
\tablenotetext{b}{Values and errors derived from Gaussian fit.}
\tablenotetext{c}{Ratio of transition intensity to that of 
\co \jone, convolved to the same beam, in \Kkmpers. To obtain ratios
for intensities in \intunits\ multiply by $(\nu/\nu_{10})^3$.}
\tablenotetext{d}{Errors are $\pm1\sigma$, and include 
10\% $1\sigma$ calibration uncertainty in both lines.}
\tablenotetext{e}{[CII]/CO \jone\ line intensity ratio in a 4' beam, 
in \intunits.}
\tablenotetext{f}{Measured in a HPBW=4\arcmin\ beam.}
\enddata
\label{tabsubmm}
\end{deluxetable}

\newpage
\begin{deluxetable}{lrrrcrc}
\tablewidth{0pc}
\small
\tablecaption{FIR Parameters for the N159/N160 Complex}
\tablehead{
Identifier & \multicolumn{1}{c}{S$_{60}$ \tablenotemark{a}} & \multicolumn{1}{c}{S$_{100}$ \tablenotemark{a}} & 
\multicolumn{1}{c}{$\frac{S_{100}}{S_{60}}$} & \multicolumn{1}{c}{T$_{dust}$ \tablenotemark{b}} &
\multicolumn{1}{c}{L$_{FIR}$ \tablenotemark{c}} & \multicolumn{1}{c}{Eq. Sp. Type \tablenotemark{d}} \\
& \multicolumn{1}{c}{(Jy)} & \multicolumn{1}{c}{(Jy)} & 
& \multicolumn{1}{c}{(K)} &
\multicolumn{1}{c}{($10^6$ L$_\odot$)} & \multicolumn{1}{c}{(ZAMS)}\\
}
\startdata
N159W	& 204.9 & 671.8 & 3.28 & 30 & 3.40 & $5\times$O5V \\
N159E	& 179.6 & 393.7 & 2.19 & 34 & 1.86 & $3\times$O5V \\
N159S \tablenotemark{e}	& 6.6   & 39.5  & 6.01 & 25 & $<0.24$ & $<$O6V \\
N160	& 393.1 & 580.1 & 1.48 & 40 & 2.74 & $4\times$O5V \\
N160E	& 48.5  & 76.1  & 1.57 & 39 & 0.36 & O5.5V \\

\tablenotetext{a}{Point source flux density at 60 and 100 $\mu$m 
derived from IRAS HIRES data.}
\tablenotetext{b}{Dust temperature derived from the ratio of S$_{100}$ 
to S$_{60}$, assuming the dust emits as a graybody with 
emissivity exponent $\beta=1$.}
\tablenotetext{c}{FIR luminosity computed according to Lonsdale (1985), 
assuming a dust emissivity exponent $\beta=1$ and $D=52$ kpc.}
\tablenotetext{d}{Equivalent spectral type and multiplicity (Panagia 1973),
assuming all starlight is absorbed by dust and reemitted in the FIR.}
\tablenotetext{e}{There is no infrared source associated with N159S, 
therefore the IRAS fluxes are upper limits.}

\enddata
\label{tabfir}
\end{deluxetable}

\newpage
\begin{deluxetable}{lccrrc}
\tablewidth{0pc}
\small
\tablecaption{Radio Continuum Sources in the N159/N160 Complex}
\tablehead{
Identifier & R.A. & Dec.       &\multicolumn{1}{c}{S$_{1.42}$ \tablenotemark{a}} & \multicolumn{1}{c}{$U$ \tablenotemark{b}} & \multicolumn{1}{c}{Sp. Type \tablenotemark{c}} \\
& (B1950) & (B1950) &\multicolumn{1}{c}{(mJy)} & \multicolumn{1}{c}{(pc cm$^{-2}$)} & \multicolumn{1}{c}{(ZAMS)}\\
}
\startdata
N159-I	& $5^{\rm h}40^{\rm m}29\fs3$ & $-69$\arcdeg46\arcmin21\arcsec & 480 & 143 & $1.5\times$O5V \\
N159-II	& $5^{\rm h}40^{\rm m}05\fs2$ & $-69$\arcdeg47\arcmin29\arcsec & 440 & 139 & $1.4\times$O5V \\
N159-III& $5^{\rm h}40^{\rm m}22\fs4$ & $-69$\arcdeg46\arcmin37\arcsec & 336 & 127 & $1.3\times$O5V \\
N159S \tablenotemark{d}	& \ldots & \ldots & $<30$   & $<57$  & $<$O6V \\
N160-I	& $5^{\rm h}40^{\rm m}07\fs8$ & $-69$\arcdeg40\arcmin17\arcsec & 802 & 170 & $1.7\times$O5V \\
N160-II	& $5^{\rm h}40^{\rm m}47\fs3$ & $-69$\arcdeg41\arcmin42\arcsec & 169 & 101 & O5V \\
N160-III& $5^{\rm h}40^{\rm m}35\fs1$ & $-69$\arcdeg41\arcmin49\arcsec & 126 & 92  & O5V \\

\tablenotetext{a}{Point source flux density at 21 cm.}
\tablenotetext{b}{\hii\ region excitation parameter.}
\tablenotetext{c}{Equivalent spectral type and multiplicity, 
derived using $U$ computed by Panagia (1973).}
\tablenotetext{d}{No radio continuum detected down
to $1\sigma\approx10$ mJy.}

\enddata
\label{tabradio}
\end{deluxetable}


\newpage

\begin{references}
\reference{AB92} Abgrall, H., Le Bourlot, J., Pineau des For\^ets, G., Roueff,
E., Flower, D. R., \& Heck, L. 1992, \aap, 253, 525






\reference{BJI99} Bolatto, A. D., Jackson, J. M., \& Ingalls, J. G. 1999,
\apj, 513, 275

\reference{BO99} Bolatto, A. D., Jackson, J. M., Wilson, C. D., \& 
Moriarty-Schieven, G. 2000, \apj, 532, in print

\reference{BO85} Bouchet, P., Lequeux, J., Maurice, E., Pr\'evot, L., \&
Pr\'evot-Burnichon, M. L. 195, \aap, 149, 330



\reference{CH97} Chu, Y.-H., Kennicutt, R. C., Snowden, S. L., Smith, R. C.,
Williams, R. M., \& Bomans, D. J. 1997, \pasp, 109, 554

\reference{CHI00} Cioni, M. R., Habing, H. J., \& Israel, F. P. 2000, \aap, 
358, L9

\reference{CO88} Cohen R. S., Dame T. M., Garay G., Montani J., Rubio M., \& 
Thaddeus P. 1988, \apjl, 331, L95

\reference{CO98} Comer\'on, F., \& Claes, P. 1998, \aap, 335, L13

\reference{CO95} Cowley, A. P., Schmidtke, P. C., Anderson, A. L., 
McGrath, T. K. 1995, \pasp, 107, 145

\reference{DEM76} Davies, R. D., Elliot, K. H., \& Meaburn, J. 1976, 
\memras, 81, 89

\reference{DB98} de Boer, K. S., Braun, J. M., Vallenari, A., \& Mebold, U. 
1998, \aap, 329, L49 

\reference{DU84} Dufour, R. J. 1984, in Structure and Evolution of the 
Magellanic Clouds, ed. S. van der Bergh \& 
K. S. de Boer (Dordrecht:Kluwer), 353






\reference{FC86} Franco, J., \& Cox, D. P. 1986, \pasp, 98, 1076

\reference{GP00} Gerin, M., \& Phillips, T. G. 2000, \apj, in print

\reference{HA68} Habing, H. J. 1967, \bain, 19, 421

\reference{HE99} Heikkil\"a, A., Johansson, L. E. B., \& Olofsson, H. 1999,
\aap, 344, 817

\reference{HE56} Henize, K. G. 1956, \apjs, 2, 315







\reference{I97} Ingalls, J. G., Chamberlin, R. A., Bania, T. M., Jackson, 
J. M., Lane, A. P., \& Stark, A. A. 1997, \apj, 479, 296




\reference{IS93} Israel, F.P. et al. 1993, \aap, 276, 25

\reference{IS96} Israel, F. P., Maloney, P. R., Geis, N., Hermann, F., 
Madden, S.C., Poglitsch, A., \& Stacey, G. J. 1996, \apj, 465, 738

\reference{IS97} Israel, F. P. 1997, \aap, 328, 471

\reference{JK99} Jackson, J. M., \& Kraemer, K. E. 1999, \apj, 512, 250

\reference{JO94} Johansson, L. E. B., Olofsson, H., Hjalmarson, \AA., Gredel, R., \& Black, J. H. 1994, \aap, 292, 371 

\reference{JO98} Johansson, L. E. B., et al. 1998, \aap, 331, 857

\reference{KA99} Kaufman, M. J., Wolfire, M. G., Hollenbach, D. J., \& Luhman,
M. L. 1999, \apj, 527, 795



\reference{KI98} Kim, S., Staveley-Smith, L., Dopita, M. A., Freeman, K. C.,
Sault, R. J., Kesteven, M. J., \& McConnell, D. 1998, \apj, 503, 674



\reference{KO82} Koornneef, J. 1982, \aap, 107, 247

\reference{KU97} Kutner, M. L., et al. 1997, \aaps, 122, 255





\reference{LE79} Lequeux, J., Peimbert, M., Rayo, J. F., Serrano, A., 
\& Torres-Peimbert, S. 1979, \aap, 80, 155


\reference{LD96} Lepp, S., \& Dalgarno, A. 1996, \aap, 306, L21


\reference{LF98} Lisenfeld, U., \& Ferrara, A. 1998, \apj, 496, 145


\reference{MA97} Madden, S. C., Poglitsch, A., Geis, N., Stacey, G. J., \&
Townes, C. H. 1997, \apj, 483, 200






\reference{MF84} Mathewson, D. S., \& Ford, V. L. 1984, in Structure and
Evolution of the Magellanic Clouds, ed. S. van den Bergh
\& K. S. de Boer (Dordrecht:Reidel), 125 


\reference{MT93} Meixner, M., \& Tielens, A. G. G. M. 1993, \apj, 405, 216

\reference{MT95} Meixner, M., \& Tielens, A. G. G. M. 1995, \apj, 446, 907




\reference{MO96} Mochizuki, K. et al. 1994, \apjl, 430, L37 






\reference{PA98} Pak, S., Jaffe, D. T., van Dishoeck, E. F., Johansson, 
L. E. B., \& Booth, R. S. 1998, \apj, 498, 735

\reference{PA73} Panagia, N. 1973, \aj, 78, 929

\reference{PA97} Panagia, N., Gilmozzi, R., \& Kirshner, R. P. 1997, in ASP
Conf. Ser., SN 1987A: Ten Years After, ed. M. Phillips \& N. Suntzeff 
(San Francisco:ASP), in print 



\reference{PO95} Poglitsch, A., Krabbe, A., Madden, S. C., Nikola, T., Geis,
N., Johansson, L. E. B., Stacey, G. J., \& Sternberg, A. 1995, \apj, 454, 293


\reference{RLB93} Rubio, M., Lequeux, J., \& Boulanger, F. 1993, \aap, 271, 9




\reference{SC99} Schmidtke, P. C., Ponder, A. L., \& Cowley, A. P. 1999, 
\aj, 117, 1292








\reference{SP78} Spitzer, L. 1978, Physical Processes in the Interstellar
Medium (New York:John Wiley \& Sons), 47



\reference{ST97} Stark, A. A., Bolatto, A. D., Chamberlin, R. A., 
Lane, A. P., Bania, T. M., Jackson, J. M., \& Lo, K.-Y. 1997a, \apjl, 480, L59

\reference{ST97} Stark, A. A., Chamberlin, R. A., Cheng, J., Ingalls, J. G.,
\& Wright, G. 1997b, Rev. Sci. Inst., 68 (5), 2200












\reference{WA96} Warin, S., Benayoun, J. J., \& Viala, Y. P. 1996, \aap, 308, 535




\reference{WI97} Wilson, C. D. 1997, \apjl, 487, L49












\end{references}
\end{document}